\documentclass[twocolumn]{aastex631}

\usepackage{rotating}
\usepackage{hyperref, amsmath}
\usepackage{tikz}
\def\checkmark{\tikz\fill[scale=0.4](0,.35) -- (.25,0) -- (1,.7) -- (.25,.15) -- cycle;} 
\usepackage{makecell}
\usepackage{multirow}

\renewcommand{\edit}[1]{}

\begin{document}

\title{Procedures for Constraining Robotic Fiber Positioning for Highly Multiplexed Spectroscopic Surveys: The Case of FPS for SDSS-V}

\correspondingauthor{Ilija Medan}
\email{ilija.medan@vanderbilt.edu}

\author[0000-0003-3410-5794]{Ilija Medan}
\affiliation{Department of Physics and Astronomy,
	Vanderbilt University,
	Nashville, TN 37235, USA}
    
\author[0000-0002-4459-9233]{Tom Dwelly}
\affiliation{Max-Planck-Institut f\"ur extraterrestrische Physik, Gie{\ss}enbachstra{\ss}e 1, 85748 Garching, Germany}

\author[0000-0001-6914-7797]{Kevin R. Covey}
\affiliation{Department of Physics and Astronomy, Western Washington University, 516 High Street, Bellingham, WA 98225, USA}

\author[0000-0003-3769-8812]{Eleonora Zari}
\affiliation{Dipartimento di Fisica e Astronomia, Universit{\`a} degli Studi di Firenze, Via G. Sansone 1, I-50019, Sesto F.no (Firenze), Italy}
\affiliation{Max-Planck-Institut f\"ur Astronomie, Konigstuhl 17, D-69117,
Heidelberg, Germany}

\author[0000-0003-1641-6222]{Michael R.~Blanton}
\affiliation{Center for Cosmology and Particle Physics,
Department of Physics,
New York University, 
726~Broadway~Rm.~1005,
New York, NY 10003, USA}

\author[0000-0001-5926-4471]{Joleen K. Carlberg}
\affiliation{Space Telescope Science Institute,
3700 San Martin Dr., Baltimore, MD 21218}

\author[0000-0003-0346-6722]{S. Drew Chojnowski}
\affiliation{NASA Ames Research Center, Moffett Field, CA 94035, USA}

\author[0000-0002-4863-8842]{Alexander Ji}
\affiliation{Department of Astronomy \& Astrophysics, University of Chicago, 5640 S Ellis Avenue, Chicago, IL 60637, USA}
\affiliation{Kavli Institute for Cosmological Physics, University of Chicago, Chicago, IL 60637, USA}

\author[0000-0003-1659-7035]{Yue Shen}
\affiliation{Department of Astronomy, University of Illinois at Urbana-Champaign, Urbana, IL 61801, USA}
\affiliation{National Center for Supercomputing Applications, University of Illinois at Urbana-Champaign, Urbana, IL 61801, USA}

\author[0009-0000-4049-5851]{John Donor}
\affiliation{Department of Physics and Astronomy, Texas Christian University, TCU Box 298840 Fort Worth, TX 76129, USA}

\author[0000-0003-2486-3858]{Jos\'{e} S\'{a}nchez-Gallego}
\affiliation{Department of Astronomy, Box 351580, University of Washington, Seattle, WA 98195, USA}

\author[0000-0002-6770-2627]{Sean Morrison}
\affiliation{Department of Astronomy, University of Illinois at Urbana-Champaign, Urbana, IL 61801, USA}

\author[0000-0002-9790-6313]{H\'ector J. Ibarra-Medel}
\affiliation{Instituto de Astronom\'ia, Universidad Nacional Aut\'onoma de M\'exico, \\
A.P. 70-264, Ciudad de M\'exico, CDMX 04510, Mexico}

\author[0000-0002-4454-1920]{Conor Sayres}
\affiliation{Department of Astronomy, Box 351580, University of Washington, Seattle, WA 98195, USA}

\author[0000-0002-3481-9052]{Keivan G. Stassun}
\affiliation{Department of Physics and Astronomy,
	Vanderbilt University,
	Nashville, TN 37235, USA}

\begin{abstract}
One crucial aspect of planning any large scale astronomical survey is constructing an observing strategy that maximizes reduced data quality. This is especially important for surveys that are rather heterogeneous and broad-ranging in their science goals. The Sloan Digital Sky Survey V (SDSS-V), which now utilizes the Focal Plane System (FPS) to robotically place fibers that feed the spectrographs, certainly meets these criteria. The addition of the FPS facilities an increase in survey efficiency, number of targets and target diversity, but also means the positions of fibers must be constrained to allow for simultaneous observations of sometimes competing programs. The constraints on the positions of the fibers are clearly driven by properties of the science targets e.g., the type of target, brightness of the target, position of the target relative to others in the field, etc. The parameters used to describe these constraints will also depend on the intended science goal of the observation, which will vary with the types of objects requested for the particular observation and the planned sky conditions for the observation. In this work, we detail the SDSS-V data collection scenarios, which consist of sets of parameters that serve as the framework for constraining fiber placements. The numerical values of these parameters were set based on either past experiences or from a series of new tests, which we describe in detail here. These parameters allow a survey like SDSS-V to be algorithmically planned to maximize the science output, while guaranteeing data quality throughout its operation.
\end{abstract}


\section{Introduction} \label{sec:intro}

The Sloan Digital Sky Survey V \citep[SDSS-V;][]{koll2025} has transformed the way that SDSS 
performs wide-field multi-object spectroscopy. Until 2021, SDSS used a fiber plug-plate
system to observe each region of sky; this system required each plug-plate to be planned
months prior to observation, to be plugged by hand the day before, and incurred a long (20 minute)
overhead for changes between plates, and with the available hardware as of SDSS-IV,
limited the number of different plug-plates possible to observe each night. 
These restrictions limited the flexibility of SDSS,
preventing it from efficiently covering larger areas of sky, optimizing strategy for a wide range of target brightnesses, or changing observing plans
rapidly.

For SDSS-V, the main hardware advancement is the focal plane system \citep[FPS;][]{FPS}, which utilizes robotic fiber positioners that enable efficient changes between telescope pointings. Because of the survey efficiency that comes with the new FPS, this allows for the selection of targets that are greater in number and variety than previous iterations of SDSS.
This system is also capable of \textit{simultaneously} observing targets with more than one instrument and across many different science programs. With these added capabilities, SDSS-V will be able to conduct a comprehensive all-sky survey that provides both optical and IR
spectra. This is possible as the FPS has 
the ability to observe at least 20--30 different fiber
configurations each night, allowing for multi-epoch visits for large regions of the sky within the length of the survey.
Such a panoptic survey is the first optical plus infrared survey of multi-object spectroscopy (MOS) that will be done for millions of sources spread across the entire sky.




These added capabilities result in a greater complexity in survey planning though. Indeed, for a survey of this size, all fiber placements must be planned prior to observation. For SDSS-V, we have built up a suite of software that allows us to fully plan the survey prior to observations. To do this algorithmically though, the constraints must be clearly defined for these pieces of software. These constraints broadly fall into three categories. First are the physical constraints of our system, such as which robots can physically reach which locations in the field of view, the altitude limits of the telescope, etc. The second are what science targets should be assigned to fibers to best accomplish the science goals of the overall survey. This will involve optimizing the assignments to maximize science output based on some metrics. Finally, the position and assignment of fibers must be constrained to ensure the resulting reduced data meets the quality needed for the varying science goals of the targets within each observation. The first two points are concepts discussed in other papers and will not be covered here \citep[see][]{Sayres2021, blanton2025}. This work specifically focuses on the latter case of constraining fiber placement to ensure the quality of the reduced data in a way that respects the science goals of the varying, and sometimes competing, programs within an observation.

The science programs of SDSS-V will range from our closest stellar neighbors to distant quasars. With such a variety, this means that targets span a much wider dynamic range in
flux than previous SDSS surveys, such that within the same observation we will often be attempting to schedule targets that compete for fibers, and that are not simultaneously observable. So, we must design a system that prevents the observation of mutually exclusive targets. Resolving the tensions between programs involves some level of compromise or amelioration, and so it becomes essential to set up new processes in order to regularize the constraints placed on the production and execution of survey \textit{Designs}. With such a large number of targets to be included in observations, it is necessary that this process is algorithmic in nature, so the decisions can be made by our automated survey planning tool. It is also important that these constraints are as minimal as possible, such that the science goals are still achieved while maximizing the number of observable targets within a single observation.

In this work, we detail a framework for making these types of decisions. Largely, these observational constraints will be broken into two broad categories; limits set on when an observation can be made and limits on how fibers can be assigned within an observation. For the latter, this will be a function of e.g., the type of target, the brightness of the target, the position of the target relative to others in the field, etc. Each of these constraints will be defined by some parametrization, such that our survey planning tool can algorithmically make decisions about targeting. This framework will be flexible enough to allow for multiple data collection scenarios, where the end science goals will differ based on the types of targets prioritized within that observation.

The decisions and lessons learned here will be crucial not only in the understanding of the survey planning for SDSS-V. This detailed information will also be crucial for users of the SDSS-V data. Specifically, to do detailed modeling of the SDSS-V selection function, these constraints on fiber assignments will act as strict limits on what objects are observable throughout the survey. In addition, the framework laid out here can be applied to other 
spectroscopic surveys that utilize similar robotically controlled 
systems like the FPS \citep[e.g. SUBARU PFS, DESI, WEAVE, 4MOST;][]{subaru, desi, weave, 4most}.

The plan of this paper is as follows. Section~\ref{sec:sdss5} provides a high-level overview of the SDSS-V survey science goals, the specific role of the FPS, and definitions of key terms to be used throughout the paper. Section~\ref{sec:params} describes in detail the various parameters involved in constraining the design of a given set of FPS observations, including both those that pertain to the conditions on sky and those pertaining to a specific telescope pointing. Section~\ref{sec:bright} presents the special considerations involved in the case of bright stars, whether they are being intentionally targeted or avoided. Section~\ref{sec:implement} briefly summarizes the software used to implement all of these \textit{Design} constraints and considerations. Finally, Section~\ref{sec:summary} concludes with a summary.

\section{The SDSS-V Survey's Utilization of the Fiber Positioning System}\label{sec:sdss5}

\subsection{Overall Survey Science Objectives and Targeting Overview}\label{sec:targeting}

SDSS-V is an all-sky, multi-epoch spectroscopic survey that will observe over six million objects \citep{sdssV, koll2025}. Its three scientific programs are referred to as ``Mappers". The Milky Way Mapper (MWM) aims to map the stellar populations and chemodynamics of the Milky Way to understand its evolution, and probe stellar physics and system architecture by observing a variety of stars in the Milky Way and Magellanic Clouds. The Black Hole Mapper (BHM) will  study the physics of black hole growth and provide optical spectra for counterparts to extragalactic X-ray selected sources from eROSITA \citep{Merloni2012, Predehl2021}. Finally, the Local Volume Mapper (LVM) will examine gas emission, star formation and stellar/interstellar energy exchange processes in the Milky Way and the Local Group.  LVM is operationally distinct from MWM and BHM, in the sense that it conducts observations using ultra-wide-ﬁeld optical IFU spectroscopy with a different set of telescopes than MWM and BHM. Due to these differences, the observing constraints of LVM will not be discussed in the present work.

Both MWM and BHM utilize the SDSS-V multi-object spectroscopic (MOS) databases \citep[see][for a description]{sdssDR18} and targeting software. This infrastructure is set up to collect observations at both the Sloan Foundation 2.5-m telescope at Apache Point Observatory (APO) in New Mexico, USA \citep{APO} and the du Pont 100-inch telescope at LCO in Chile \citep{LCO}. \edit1{For SDSS-V, the sky is split between these two sites, where we do not design overlapping \textit{Fields} between the two observatories.} Each observatory is equipped with an optical \citep[BOSS;][]{smee13a} and an IR \citep[APOGEE;][]{wilson19a} fiber-fed spectrograph. The main hardware advancement of SDSS-V is the focal plane system \citep[FPS;][]{FPS}, which utilizes 500 robotic fiber positioners that enable efficient robotic re-configuration of the on-sky locations of fibers. Because of the survey efficiency and flexibility that comes with the new FPS, this allows for MWM and BHM to select targets that are greater in number and variety than previous iterations of SDSS. Also, in past SDSS-I--IV plate observations, the typical time per pointing was 1--3 hours due to the overheads of replacing the cartridges holding the plates and the desire to observe overall fainter targets. With the FPS for SDSS-V, we can reduce exposure times in fields with many bright targets to one single, 15 minute exposure for each pointing.

A full description of the targeting for SDSS-V can be found in the DR18 paper \citep{sdssDR18}, but we provide a summary here to demonstrate the variety of objects targeted in SDSS-V, which greatly increases the complexity of the survey constraints considered. For MWM, there are four main focuses that guide target selection. First is Galactic archaeology, which will reconstruct the history of the Milky Way by analyzing the number, masses, composition, ages and motions of stars. Here luminous red giants throughout the Galactic disk will serve as the tracers of this history. Second, MWM will study the young Galaxy missed by the previously used luminous red giants. Here the early phases of low-mass stellar evolution and the life cycles of massive stars will be examined, by systematically targeting both low-mass, young stars and hot stars. Third, MWM will study the high-energy components of the Milky Way, which include targets like compact objects, young stellar objects and flaring stars. Finally, MWM will generally study a wide variety of stellar physics and stellar systems. This includes a census of stars in the Solar Neighborhood, understanding stellar structure via astreoseismoilogy, and observing a variety of binaries, particularly compact binaries. This breadth of science requires MWM to target stars across the HR diagram. 

BHM's science goal on the other hand is to study the growth of supermassive black holes at the centers of galaxies. It will accomplish this through time-domain spectroscopy of quasars, as well as single-epoch spectroscopy of X-ray selected active galactic nuclei and galaxy clusters. While the number of categories of objects may be smaller in BHM compared to MWM, the targets in BHM will require a broad range of timescales and cadence of observation, as different aspects of quasar physics result in variability on very different timescales (days to years). The BHM time-domain science goals set demanding requirements on the spectrophotometric accuracy of the delivered optical spectra, placing constraints on the calibration strategy. Additionally, BHM targets will push the overall faint end of the distribution of targets in the survey, creating a larger overall dynamic range in flux for all of the SDSS-V targets.

\subsection{Definition of Terms}

For the remainder of this work, we will make use of specific terminology as a shorthand to refer to various aspects of the FPS \textit{Design} procedures. In Table \ref{tab:term-definitions} we provide a summary of these terms and their definitions for convenience. The reader can reference this table when such terms appear throughout the following sections.

\begin{table*}[!t]
  \centering
  \caption{Definition of Terms} 
  \begin{tabular}{|l | l |p{0.5\linewidth}|}
\hline
       \textbf{Category} & \textbf{Term} & \textbf{Definition} \\
        \hline
        \hline
        Data Collection Scenarios & Bright Time & MWM driven scenario used in crowded field observations that can tolerate high lunation \\
        \cline{2-3}
                                  & Dark Plane & MWM driven scenario used for spectroscopic observations of moderately faint targets observed in crowded fields (i.e., the MW plane). \\
        \cline{2-3}
                                  & Dark Faint & BHM driven scenario used for spectroscopic observations of faint targets in uncrowded fields, requiring the darkest skies.\\
        \cline{2-3}
                                  & Dark Monitoring & BHM driven scenario used for spectroscopic observations that must attain a high level of spectrophotometric accuracy and good signal to noise for moderately faint targets \\
       \cline{2-3}
                                  & Dark RM & Scenario specifically designed for the BHM reverberation mapping program.  \\
\hline
\hline
        FPS Observation Constraints & \texttt{obsmode}    & Specifications of the desired sky conditions at the time of observation. \\
        \cline{2-3}
                                    & \texttt{designmode} & Specifications of spectrograph specific calibration requirements and fiber assignment restrictions. \\
        \hline
        \hline
        FPS \textit{Design} Units & \textit{Field}  & A location on the sky (center and rotation angle of the telescope field of view) for which there are one or more \textit{Designs} whose fiber assignments will be determined in coordination with each other \\
        \cline{2-3}
                                  & \textit{Design} & A definition of a planned observation within a \textit{Field} with a defined set of fiber assignments. \\
        \hline
        \hline
        SDSS-V Software Products  & \texttt{kaiju}        & Generates collision free paths of robots in the FPS. \\
        \cline{2-3}
                                  & \texttt{coordio}      & Performs coordinate conversions between internal FPS systems and on sky positions. \\
        \cline{2-3}
                                  & \texttt{robostrategy} & Survey planning software that accounts for all \textit{Field} and \textit{Design} level constraints. \\
        \cline{2-3}
                                  & \texttt{mugatu}       & \textit{Design} validation software that checks and calculates metrics based on the \texttt{designmode} parameters for a given \textit{Design}. \\
      \hline
      \end{tabular}
  \label{tab:term-definitions}
\end{table*}

\subsection{FPS Data Collection Scenarios}\label{sec:obs_mode}

As discussed in Section \ref{sec:targeting}, MWM and BHM targets will cover a large range of types of astrophysical objects, 
magnitudes, object densities per sky location, and observing cadence 
requirements. To efficiently plan a survey with such diversity, a set of constraints must be defined to facilitate algorithmic planning of observations that guarantee sufficiently good science outcomes for all of its defined goals. \edit1{Their is a balance that must be struck here though. First, what may be a ``good" science outcome varies from program to program. So, the set of parameters used to constrain assignments must be flexible enough to accommodate this. Second, the survey is finite. So, while observations could be done continuously to achieve the perfect science outcomes, this does not fall within the bounds of a limited survey. This means that the set of constraints must be flexible enough to fit within such limits. With these considerations, we divide survey requirements} into various SDSS-V ``data collection scenarios", such that each program can select a data collection scenario that best fits their science needs. These goals are set by the main science targets within a \textit{Field}, and not \edit1{by the} additional open fiber or filler targets that are added at the end of the \textit{Design} making process to fill unused fibers.
The needs of the BHM and MWM science goals required the definition of five distinct scenarios: 
 Bright Time, 
Dark Plane, Dark Monitoring, Dark Reverberation Mapping (RM), and Dark Faint. Each of these scenarios is associated with a set of carefully parameterized \textit{Design} constraints that best address the main science drivers that utilize the 
scenario. 
They are defined as follows: 
\begin{itemize}
\item Bright Time is optimized for APOGEE 
and BOSS spectroscopic observations during sky conditions with bright lunation. It is driven by MWM science goals that require stellar parameters, individual chemical abundances, and radial velocities.
The 
parameters of this scenario are set to ensure reliable sky subtraction and relative flux 
calibration, with specific consideration that data will be collected in 
crowded fields and in high lunar illumination conditions. 
\item Dark Plane is optimized for BOSS observations during dark lunation in pursuit of MWM science goals, particularly for obtaining stellar velocities, emission
 line fluxes and spectral classifications of (optically) faint 
point-like targets. Like Bright Time, requirements are set to ensure 
reliable sky subtraction and relative flux calibration, especially in 
crowded fields. 
\item Dark Faint on the other hand is driven primarily by BHM science goals and is optimized for BOSS spectroscopy to
 measure redshifts and classify (optically) faint point-like and extended targets in
 uncrowded conditions. Again, these requirements are set to ensure the best possible 
sky subtraction and reasonably accurate relative flux calibration. 
\item Dark Monitoring 
is optimized for high spectrophotometric accuracy during BOSS observations of 
moderately faint point-like extragalactic targets (QSOs), to produce time series of calibrated spectra that can be 
compared over month-to-year baselines. 
\item Dark RM has very similar goals to Dark Monitoring, but is further optimized to ensure 
high quality and efficient observations of the BHM reverberation mapping (RM) program. 
\end{itemize}
\edit1{SDSS-V programs select one of these above data collection scenarios via a specific observing cadence, which is always associated with a scenario. During the survey planning (more details in Section \ref{sec:implement}), while multiple programs will be possible to observe within a \textit{Field}, the program that is dominant will drive the choice to assign a single data collection scenario for the entire \textit{Field}. The exception is mixed cadence \textit{Fields}, where the first set of exposures can be a dark data collection scenario and the rest Bright Time. This allows for more dark time observations within the plane.}

\edit1{In general,} RM \edit1{dominated} \textit{Fields} use the
Dark RM scenario, \textit{Fields} associated with the All-Quasar Multi-Epoch Spectroscopy
(AQMES) program use Dark Monitoring, \textit{Fields} associated with the  SPectroscopic 
IDentfication of ERosita Sources (SPIDERS) program use Dark Faint, and all other
dark-time \edit1{assigned} \textit{Fields} (typically dark time at low Galactic latitude, which allows 
observations of white dwarfs) use Dark Plane. Bright Time \textit{Fields} are those that are dominated by MWM targets.

\section{FPS Observation Constraint Parameters}\label{sec:params}

Each data selection scenario can be defined by a unique set of 
constraints that fall into one of two categories: observing conditions 
and fiber assignment restrictions. For SDSS-V, these constraints are defined 
by the \texttt{obsmode} and \texttt{designmode} parameters, 
respectively. The \texttt{obsmode} parameters impact what 
classes of 
science targets can be assigned to fibers on a \textit{Design} based on the intended observing conditions and also affect when the \textit{Design} can be scheduled for observation. The \texttt{designmode}, in contrast, impacts  
how fibers can be assigned to individual targets within the \textit{Design}. These requirements ensure that \textit{Designs} contain the required number of calibrators, that targets meet the 
minimum 
and maximum magnitude criteria for the \textit{Design}, and that fibers are not placed near bright sources that could affect the reduction of other science targets within the \textit{Design}. In general, there is a balance between defining very strict observational and \textit{Design} constraints (which guarantee data quality) versus a desire to allow observations to be carried out under the widest possible set of conditions (increasing schedulability), and allowing heterogeneous combinations of targets to be observed simultaneously (increasing \textit{Design} flexibility). For both categories of constraints, the parameters must be tuned to strike a balance between these two extremes, where these decisions are informed by past experience and specifically designed tests to assess their impact. \edit1{Additionally, as will be shown throughout this work, we find in specific cases that 1) it is not always possible to enforce these requirements, or 2) not all requirements are strictly necessary to, on average, produce data at a quality needed for the survey science requirements, especially at the expense of schedulability and \textit{Design} flexibility. Such instances will be explicitly discussed in the remainder of this paper and we will demonstrate that not meeting requirements in these limited cases does not impact the overall quality of the survey.}



\subsection{Determining \texttt{obsmode} Constraints}\label{sec:obsmode}

The first set of parameters used to constrain SDSS-V observations are the \texttt{obsmode} parameters; specification of the desired sky conditions at the time of observation. These \texttt{obsmode} parameters are used by the scheduler to know when it is acceptable to observe a \textit{Design} in a \textit{Field} \citep{donor2024}. In practice, these parameters can change from \textit{Design} to \textit{Design} within a given \textit{Field}, so in order to constrain the observation of a given \textit{Design}, we consider the following \texttt{obsmode} parameters:
\begin{itemize} 
    \item \texttt{min\_moon\_sep}: minimum angular separation, in decimal degrees, between the Moon’s position at the time of observation and the center of the FPS Field of View.
    \item \texttt{min\_deltaV\_KS91}: minimum acceptable difference between the local expected V-band sky brightness at time of observation and that predicted for a zenith field at new moon. The scheduler predicts this sky brightness difference at the time of observation using the $\Delta V$ framework developed by \cite{Krisciunas_1991}. Brighter skies produce more negative $\Delta V$ values (ie, a $\Delta V = -2.3$ sky is brighter than a $\Delta V = -1.0$ sky), so \texttt{min\_deltaV\_KS91} limits will identify the minimum value under which observations are allowed; darker skies will have $\Delta V$ values that are larger than this limit (i.e. they will be closer to zero).
    \item \texttt{min\_twilight\_ang}: minimum acceptable solar angular distance below the horizon at the time of observation. A \textit{Design} that can only be observed beyond $15^\circ$ twilight would have \texttt{min\_twilight\_ang} $= +15^\circ$; a \textit{Design} that can be observed earlier, after $8^\circ$ twilight, would have \texttt{min\_twilight\_ang} $= +8^\circ$.
    \item \texttt{max\_airmass}: maximum acceptable airmass for observations in this \texttt{obsmode}. Observing at higher airmasses allows greater flexibility when planning and executing the survey \citep{blanton2025}, so maximum airmass limits are chosen to ensure data quality with the ``lead" instrument (BOSS or APOGEE) for each \texttt{obsmode}. This is especially important for BOSS led fields, as optical observations are more affected by high airmass observations.
\end{itemize}
Table \ref{tab:obsmode-moonsky-definitions} shows the parameters for all possible data collection scenarios that could be used by a given \textit{Field}.

While both spectrographs are used for every exposure, bright time science is primarily driven by targets using the APOGEE spectrograph, and dark time science is primarily driven by BOSS targets. Requirements for each data collection scenario reflect the spectrograph driving that observing mode. 

During Bright Time, all \texttt{obsmode} parameters except \texttt{min\_deltaV\_KS91} were adopted from previous SDSS surveys \citep{apogee_overview}. The value for \texttt{min\_deltaV\_KS91} is the result of finding the limit where the APOGEE data reduction pipeline failed and then choosing a value larger than this. This was done to maximize possible sky conditions that a Bright Time \textit{Design} could be observed in.

Dark time requirements are driven by the optical BOSS spectrograph. As in bright time, the lessons of previous SDSS surveys \citep[e.g.,][]{boss, eboss}{}{} were applied to SDSS-V.  Different science drivers set the desired outcome for the different dark time data collection scenarios, however, and result in some variations between particular \texttt{obsmode} parameters.  In the BHM-led dark time scenarios (e.g., Dark Monitoring, Dark Faint and Dark RM), appropriate values for the \texttt{min\_deltaV\_KS91} parameter were determined by examining regions in the moon illumination fraction vs.~moon distance plane where good quality observations have been collected in archival
SDSS BOSS/eBOSS datasets. 
The Dark Faint data collection scenario is typically used to observe the optically faintest objects targeted by SDSS-V (to $r \sim 22$). Therefore, a stricter sky brightness constraint was used for Dark Faint (with respect to other dark modes) to ensure reasonable signal to noise at the faint end. Finally, for most dark time scenarios, the airmass limit was set to 1.4, primarily to reduce the impact of differential atmospheric refraction (based on experience from past iterations of SDSS). 
\edit1{This relatively strict constraint ensures good BOSS observing efficiency/quality, whilst also allowing access to i) the well known deep fields targeted by the reverberation mapping program, ii) the entire footprint covered by previous SDSS optical spectroscopic observations of quasars, and iii) nearly all of the high Galactic latitude sky suitable for spectroscopic follow up of X-ray selected extragalactic populations.}
However, Dark Plane extends this limit to 1.6 to improve schedulability of \textit{Designs}.

\edit1{We note that none of the above parameters constrain stochastic observing conditions. While on-sky conditions such as seeing and transparency can have a significant effect on observing efficiency, in practice these conditions can vary on short time scales and are often unpredictable, making it impossible to account for these conditions when scheduling observations. During the night, an estimate of the signal-to-noise ratio (SNR) is calculated at a fiducial magnitude for each observation. In situations where conditions are less favorable, the measured SNR will be below the requirement, and another exposure is taken.}

\begin{table}
  \footnotesize
  \centering
  \caption{Values of \texttt{obsmode} for the different data collection scenarios.} 
  \begin{tabular}{ |l|c|c|c|c|c| }
       \cline{2-6}
       \multicolumn{1}{c|}{} & \multicolumn{5}{c|}{\textbf{\textit{Data Collection Scenario}}} \\
       \hline
       \multirow{2}*{\texttt{obsmode}}& \bfseries{Bright} & \bfseries{Dark} & \bfseries{Dark} & \bfseries{Dark} & \bfseries{Dark} \\
         & \bfseries{Time} & \bfseries{Plane} & \bfseries{Monit.$^a$} & \bfseries{RM$^a$} & \bfseries{Faint} \\
      \hline
      \hline 
      \texttt{min\_lunar\_sep} & 15.0$^\circ$ & 35.0$^\circ$ & 35.0$^\circ$ & 35.0$^\circ$ & 35.0$^\circ$ \\
      \hline
      \texttt{min\_deltaV\_KS91} & -3.0 & -1.5 & -1.5 & -1.5 & -0.5 \\
      \hline
      \texttt{min\_twilight\_ang} & 8.0$^\circ$ & 15.0$^\circ$ & 15.0$^\circ$ & 15.0$^\circ$ & 15.0$^\circ$ \\ 
      \hline      
      \texttt{max\_airmass} & 2.0 & 1.6 & 1.4 & 1.4 & 1.4 \\
      \hline
      \end{tabular}
  \label{tab:obsmode-moonsky-definitions}

\footnotesize{$^a$ \textit{Design} completion criteria also differs between \texttt{obsmode}, so while Dark Monitoring and Dark RM have the same observing criteria, they have different completion requirements. Completion criteria is discussed in \citet{donor2024}}\\
\end{table}

\subsection{Determining \texttt{designmode} Constraints}\label{sec:desmode}


The \texttt{designmode} parameters focus on spectrograph-specific calibration requirements and fiber assignment restrictions. These parameters are always implemented at the \textit{Design} level and serve as the main constraints for fiber assignments at the survey planning stage. The list of \texttt{designmode} parameters were conceived to ensure that the set of observations in each \textit{Design} would meet the science goals of each data collection scenario, as outlined in Section \ref{sec:obs_mode}. These parameters control various constraints on the skies, standards and science targets for each spectrograph for a \textit{Design} in a given data collection scenario, which in turn constrains the possible fiber assignments for a \textit{Design}.

\edit1{We note that SDSS multi-object spectroscopic observations are calibrated via the simultaneous observation of appropriate calibration targets (blank sky locations, and standard/telluric stars), distributed within the field of view. For further information regarding the selection of calibration targets in SDSS-V we refer the reader to \citet{sdssDR18}. The usage of calibration targets within the BOSS and APOGEE data reduction pipelines is described by \citet{Stoughton2002,Bolton2012} and \citet{Nidever2015} respectively.}

The \texttt{designmode} parameters are as follows:
\begin{itemize}  
    \item \texttt{skies\_min:} The minimum number of fibers that must be assigned to sky locations in the \textit{Design}.
    \item \texttt{skies\_FOV:} Parameter used to assess the distribution of fibers assigned to sky across the field of view. 
    The metric is computed as follows: a) measure the focal plane distance (in mm) between each science fiber ($i$) and its $k$th nearest sky fiber ($r_{k,i}$), b) compute the distance $r_{p}$ for which $p$ percent of fibers have $r_{k,i} < r_{p}$.
    If $r_{p}$ is less than some desired distance, $d$ (in mm), then the skies are considered to be well distributed in the field of view. The \texttt{skies\_FOV} parameter is expressed as an array of $[k, p, d]$. See section \ref{sec:calib_dist} for further discussion.
    \item \texttt{stds\_min:} The minimum number of fibers that must be assigned to spectrophotometric standard stars in the \textit{Design}.
    \item \texttt{stds\_mags\_min:} The minimum (bright) magnitude limit for standards assigned in the \textit{Design}. The parameter is expressed as an array with magnitudes: $[g, r, i, z, B_P, G, R_P, J, H, K]$. If a value is null ($\emptyset$), then that magnitude is not checked when assigning a standard. A standard star can only be assigned if all magnitudes it has a measurement for are greater than the minimum values listed for this parameter.
    \item \texttt{stds\_mags\_max:} The maximum (faint) magnitude limit for standards assigned in the \textit{Design}. The parameter is expressed as an array with magnitudes: $[g, r, i, z, B_P, G, R_P, J, H, K]$. If a value is null ($\emptyset$), then that magnitude is not checked when assigning a standard. A standard star can only be assigned if all magnitudes it has a measurement for are less than the maximum values listed for this parameter.
    \item \texttt{stds\_FOV:} Parameter used to assess the the distribution of fibers assigned to standard stars across the field of view. This parameter is of the same form as \texttt{skies\_FOV}.
    \item \texttt{bright\_limit\_targets\_min:} The minimum (bright) magnitude limit for science targets assigned in the \textit{Design}. The parameter is expressed and checked the same way as \texttt{stds\_mags\_min}.
    \item \texttt{bright\_limit\_targets\_max:} The maximum (faint) magnitude limit for science targets assigned in the \textit{Design}. The parameter is expressed and checked the same way as \texttt{stds\_mags\_max}.
\end{itemize}
We set these parameters separately for BOSS and APOGEE fiber assignments, and for each of the data collection scenarios listed in Section \ref{sec:obs_mode}. The values of these \texttt{designmode} parameters for each spectrograph and scenario are listed in Table \ref{tab:targetdb-designmode-values}. In the following sections, we discuss in more detail how these metrics and parameter values were chosen.

\begin{sidewaystable*}
\vspace*{-6cm}
  \footnotesize
  \centering
  \caption{Values of \texttt{designmode} for the different data collection scenarios.} 
  \begin{tabular}{|l|l|c|c|c|c|c| }
       \cline{3-7}
       \multicolumn{2}{c|}{} & \multicolumn{5}{c|}{\textbf{\textit{Data Collection Scenario}}} \\
      \hline
       \multirow{2}*{\bfseries{Instrument}} & \multirow{2}*{\bfseries{\texttt{designmode}}}
 & \bfseries{Bright} & \bfseries{Dark} & \bfseries{Dark} & \bfseries{Dark} & \bfseries{Dark} \\
         &  & \bfseries{Time} & \bfseries{Plane} & \bfseries{Monitoring} & \bfseries{RM} & \bfseries{Faint} \\
      \hline
      \hline
      \multirow{8}*{BOSS} & \texttt{skies\_min} & 50 & 50 & 50 & 50 & 80 \\
      \cline{2-7}
       & \texttt{skies\_FOV} & [1, 95, 75] & [1, 95, 75] & [1, 95, 75] & [1, 95, 75] & [3, 95, 85] \\
      \cline{2-7}
       & \texttt{stds\_min} & 50 & 50 & 70 & 70 & 20 \\
      \cline{2-7}
       & \texttt{stds\_mags\_min} & \makecell{[$\emptyset$, $\emptyset$, $\emptyset$, $\emptyset$, 13,\\$\emptyset$, 13, $\emptyset$, $\emptyset$, $\emptyset$]} &
                              \makecell{[$\emptyset$, 15, $\emptyset$, $\emptyset$, $\emptyset$,\\$\emptyset$, $\emptyset$, $\emptyset$, $\emptyset$, $\emptyset$]} &
                              \makecell{[$\emptyset$, 16, $\emptyset$, $\emptyset$, $\emptyset$,\\$\emptyset$, $\emptyset$, $\emptyset$, $\emptyset$, $\emptyset$]} &
                              \makecell{[$\emptyset$, 16, $\emptyset$, $\emptyset$, $\emptyset$,\\$\emptyset$, $\emptyset$, $\emptyset$, $\emptyset$, $\emptyset$]} & 
                              \makecell{[$\emptyset$, 16, $\emptyset$, $\emptyset$, $\emptyset$,\\$\emptyset$, $\emptyset$, $\emptyset$, $\emptyset$, $\emptyset$]} \\
      \cline{2-7}
       & \texttt{stds\_mags\_max} &  \makecell{[$\emptyset$, $\emptyset$, $\emptyset$, $\emptyset$, $\emptyset$,\\$\emptyset$, $\emptyset$, $\emptyset$, $\emptyset$, $\emptyset$]} &
                              \makecell{[$\emptyset$, 18, $\emptyset$, $\emptyset$, $\emptyset$,\\$\emptyset$, $\emptyset$, $\emptyset$, $\emptyset$, $\emptyset$]} &
                              \makecell{[$\emptyset$, 18, $\emptyset$, $\emptyset$, $\emptyset$,\\$\emptyset$, $\emptyset$, $\emptyset$, $\emptyset$, $\emptyset$]} &
                              \makecell{[$\emptyset$, 18, $\emptyset$, $\emptyset$, $\emptyset$,\\$\emptyset$, $\emptyset$, $\emptyset$, $\emptyset$, $\emptyset$]} &
                              \makecell{[$\emptyset$, 18, $\emptyset$, $\emptyset$, $\emptyset$,\\$\emptyset$, $\emptyset$, $\emptyset$, $\emptyset$, $\emptyset$]} \\
      \cline{2-7}
       & \texttt{stds\_FOV} & [1, 95, 75] & [1, 95, 75] & [3, 95, 95] & [3, 95, 95] & [1, 95, 130] \\
      \cline{2-7}
       & \texttt{bright\_limit\_targets\_min} & \makecell{[12.7, $\emptyset$, 12.7, $\emptyset$, 13,\\13, 13, $\emptyset$, $\emptyset$, $\emptyset$]} &
                                          \makecell{[15, 15, 15, $\emptyset$, $\emptyset$,\\$\emptyset$, $\emptyset$, $\emptyset$, $\emptyset$, $\emptyset$]} &
                                          \makecell{[16, 16, 16, $\emptyset$, $\emptyset$,\\$\emptyset$, $\emptyset$, $\emptyset$, $\emptyset$, $\emptyset$]} &
                                          \makecell{[16, 16, 16, $\emptyset$, $\emptyset$,\\$\emptyset$, $\emptyset$, $\emptyset$, $\emptyset$, $\emptyset$]} &
                                          \makecell{[16, 16, 16, $\emptyset$, $\emptyset$,\\$\emptyset$, $\emptyset$, $\emptyset$, $\emptyset$, $\emptyset$]} \\
      \cline{2-7}
       & \texttt{bright\_limit\_targets\_max} & \makecell{[$\emptyset$, $\emptyset$, $\emptyset$, $\emptyset$, $\emptyset$,\\$\emptyset$, $\emptyset$, $\emptyset$, $\emptyset$, $\emptyset$]} &
      \makecell{[$\emptyset$, $\emptyset$, $\emptyset$, $\emptyset$, $\emptyset$,\\$\emptyset$, $\emptyset$, $\emptyset$, $\emptyset$, $\emptyset$]} &
      \makecell{[$\emptyset$, $\emptyset$, $\emptyset$, $\emptyset$, $\emptyset$,\\$\emptyset$, $\emptyset$, $\emptyset$, $\emptyset$, $\emptyset$]} &
      \makecell{[$\emptyset$, $\emptyset$, $\emptyset$, $\emptyset$, $\emptyset$,\\$\emptyset$, $\emptyset$, $\emptyset$, $\emptyset$, $\emptyset$]} &
      \makecell{[$\emptyset$, $\emptyset$, $\emptyset$, $\emptyset$, $\emptyset$,\\$\emptyset$, $\emptyset$, $\emptyset$, $\emptyset$, $\emptyset$]} \\
      \hline
      \hline
      \multirow{8}*{APOGEE} & \texttt{skies\_min} & 35 & 35 & 35 & 0 & 35 \\
      \cline{2-7}
       & \texttt{skies\_FOV} & [1, 95, 85] & [1, 95, 85] & [1, 95, 85] & [$\emptyset$, $\emptyset$, $\emptyset$] & [1, 95, 85] \\
      \cline{2-7}
       & \texttt{stds\_min} & 15 & 15 & 15 & 0 & 15 \\
      \cline{2-7}
       & \texttt{stds\_mags\_min} & \makecell{[$\emptyset$, $\emptyset$, $\emptyset$, $\emptyset$, $\emptyset$,\\$\emptyset$, $\emptyset$, $\emptyset$, 7, $\emptyset$]} &
                                \makecell{[$\emptyset$, $\emptyset$, $\emptyset$, $\emptyset$, $\emptyset$,\\$\emptyset$, $\emptyset$, $\emptyset$, 7, $\emptyset$]} &
                                \makecell{[$\emptyset$, $\emptyset$, $\emptyset$, $\emptyset$, $\emptyset$,\\$\emptyset$, $\emptyset$, $\emptyset$, $\emptyset$, $\emptyset$]} &
                                \makecell{[$\emptyset$, $\emptyset$, $\emptyset$, $\emptyset$, $\emptyset$,\\$\emptyset$, $\emptyset$, $\emptyset$, $\emptyset$, $\emptyset$]} &
                                \makecell{[$\emptyset$, $\emptyset$, $\emptyset$, $\emptyset$, $\emptyset$,\\$\emptyset$, $\emptyset$, $\emptyset$, 7, $\emptyset$]} \\
      \cline{2-7}
       & \texttt{stds\_mags\_max} & \makecell{[$\emptyset$, $\emptyset$, $\emptyset$, $\emptyset$, $\emptyset$,\\$\emptyset$, $\emptyset$, $\emptyset$, 13, $\emptyset$]}&
                                \makecell{[$\emptyset$, $\emptyset$, $\emptyset$, $\emptyset$, $\emptyset$,\\$\emptyset$, $\emptyset$, $\emptyset$, 13, $\emptyset$]} &
                                \makecell{[$\emptyset$, $\emptyset$, $\emptyset$, $\emptyset$, $\emptyset$,\\$\emptyset$, $\emptyset$, $\emptyset$, $\emptyset$, $\emptyset$]} &
                                \makecell{[$\emptyset$, $\emptyset$, $\emptyset$, $\emptyset$, $\emptyset$,\\$\emptyset$, $\emptyset$, $\emptyset$, $\emptyset$, $\emptyset$]} &
                                \makecell{[$\emptyset$, $\emptyset$, $\emptyset$, $\emptyset$, $\emptyset$,\\$\emptyset$, $\emptyset$, $\emptyset$, 13, $\emptyset$]} \\
      \cline{2-7}
       & \texttt{stds\_FOV} & [3, 95, 230] & [3, 95, 230] & [3, 95, 230] & [$\emptyset$, $\emptyset$, $\emptyset$] & [3, 95, 230] \\
      \cline{2-7}
       & \texttt{bright\_limit\_targets\_min} & \makecell{[$\emptyset$, $\emptyset$, $\emptyset$, $\emptyset$, $\emptyset$,\\$\emptyset$, $\emptyset$, $\emptyset$, 7, $\emptyset$]} &
      \makecell{[$\emptyset$, $\emptyset$, $\emptyset$, $\emptyset$, $\emptyset$,\\$\emptyset$, $\emptyset$, $\emptyset$, 7, $\emptyset$]} &
      \makecell{[$\emptyset$, $\emptyset$, $\emptyset$, $\emptyset$, $\emptyset$,\\$\emptyset$, $\emptyset$, $\emptyset$, 7, $\emptyset$]} &
      \makecell{[$\emptyset$, $\emptyset$, $\emptyset$, $\emptyset$, $\emptyset$,\\$\emptyset$, $\emptyset$, $\emptyset$, $\emptyset$, $\emptyset$]} &
      \makecell{[$\emptyset$, $\emptyset$, $\emptyset$, $\emptyset$, $\emptyset$,\\$\emptyset$, $\emptyset$, $\emptyset$, 7, $\emptyset$]} \\
      \cline{2-7}
       & \texttt{bright\_limit\_targets\_max} & \makecell{[$\emptyset$, $\emptyset$, $\emptyset$, $\emptyset$, $\emptyset$,\\$\emptyset$, $\emptyset$, $\emptyset$, $\emptyset$, $\emptyset$]} &
      \makecell{[$\emptyset$, $\emptyset$, $\emptyset$, $\emptyset$, $\emptyset$,\\$\emptyset$, $\emptyset$, $\emptyset$, $\emptyset$, $\emptyset$]} &
      \makecell{[$\emptyset$, $\emptyset$, $\emptyset$, $\emptyset$, $\emptyset$,\\$\emptyset$, $\emptyset$, $\emptyset$, $\emptyset$, $\emptyset$]} &
      \makecell{[$\emptyset$, $\emptyset$, $\emptyset$, $\emptyset$, $\emptyset$,\\$\emptyset$, $\emptyset$, $\emptyset$, $\emptyset$, $\emptyset$]} &
      \makecell{[$\emptyset$, $\emptyset$, $\emptyset$, $\emptyset$, $\emptyset$,\\$\emptyset$, $\emptyset$, $\emptyset$, $\emptyset$, $\emptyset$]} \\
      \hline

      \end{tabular}
     \label{tab:targetdb-designmode-values}
\end{sidewaystable*}

\subsubsection{Number of Calibrators}


Per the FPS data collection scenarios outlined in Section \ref{sec:obs_mode}, some scenarios have an emphasis on accuracy in spectrophotometry, while others prioritize the relative flux calibration of the outputs. One major component of ensuring these goals are realized for the observations of a \textit{Design} is the number of calibrators used for the reduction. In general, a larger number of standards in a \textit{Design} will result in reduced data with more accurate spectrophotometry. The question is exactly \textit{where} this line is drawn, as a balance needs to be struck between the number of fibers devoted to calibration and science targets, as maximizing the latter facilities a more rapid survey at the possible cost of poor quality data.

For the APOGEE calibrators, we relied on past experiences from APOGEE-1 \citep{Zasowski2013} and APOGEE-2 \citep{Zasowski2017}. From their commissioning tests, they found that reserving 35 fibers for sky positions was adequate for proper sky subtraction. So, all data collection scenarios that contain APOGEE science targets use this threshold. During APOGEE-2, anywhere from 15--35 standards were reserved for each observation. These telluric standards are crucial, as atmospheric H$_2$O, CO$_2$ and CH$_4$ contribute substantial absorption features to every observed IR spectrum. From this experience, it was found that at least 15 are needed to separate these lines from the stellar
and interstellar features, and perform telluric corrections. So, for all data collection scenarios 15 APOGEE standards are required.

Because about 60\% of the BOSS fibers share their robot with an 
APOGEE fiber, in \textit{Fields} with many BOSS targets, there are many
unused APOGEE fibers. Although these fibers are relatively close to
a BOSS target, the BOSS targets tend to be faint. In general, especially
outside the Galactic Plane,
the majority of the unused APOGEE fibers are pointing to perfectly good 
$H$-band sky locations. We find that essentially all \textit{Designs} $|b|>20^\circ$ 
have enough such fibers that we do not have to explicitly assign APOGEE
sky fibers. Therefore, in these \textit{Fields} we can implement a version of each 
{\tt designmode} that requires zero APOGEE skies. In versions of the survey plan to be released for DR21, we utilize such {\tt designmode}'s for \textit{Designs} in \textit{Fields} with $|b|>20^\circ$ .

For the BOSS calibrators, we do not use such a uniform approach in the parameters as has been done with APOGEE. To test the various scenarios, we utilized pre-SDSS-V plate data. Observations were made with the BOSS spectrograph during the eBOSS survey \citep{eboss}. Although these observations did not utilize the FPS, they serve as a good proxy for SDSS-V observations. Specifically, the four plate-MJDs used were: 7338--57490 (consisting of seven 15 min exposures with $<30\%$ moon illumination), 7340--58289 (consisting of ten 15 min exposures with 50\% moon illumination), 7339-57428 (consisting of four 15 min exposures) and 7338-57038 (consisting of nine 15 min exposures with $<30\%$ moon illumination). For each plate, we artificially masked some number of skies and standards to mimic some number of calibrators, and reran the pipeline to examine the resulting data. These data were then compared to the results when all calibrators in the plate were used.

\begin{figure*}[!t]
	\centering
	\includegraphics[width=0.49\textwidth]{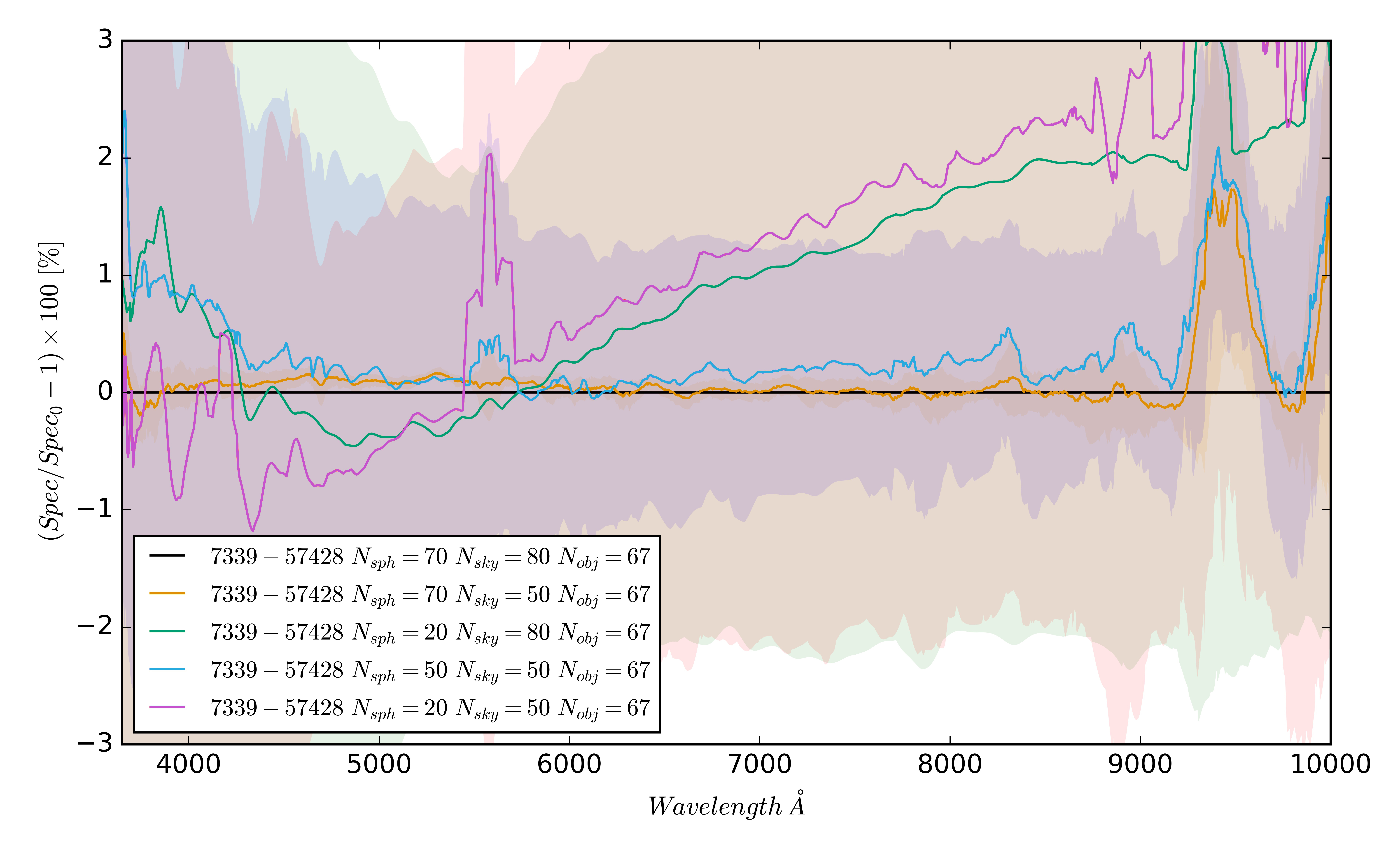}
    \includegraphics[width=0.49\textwidth]{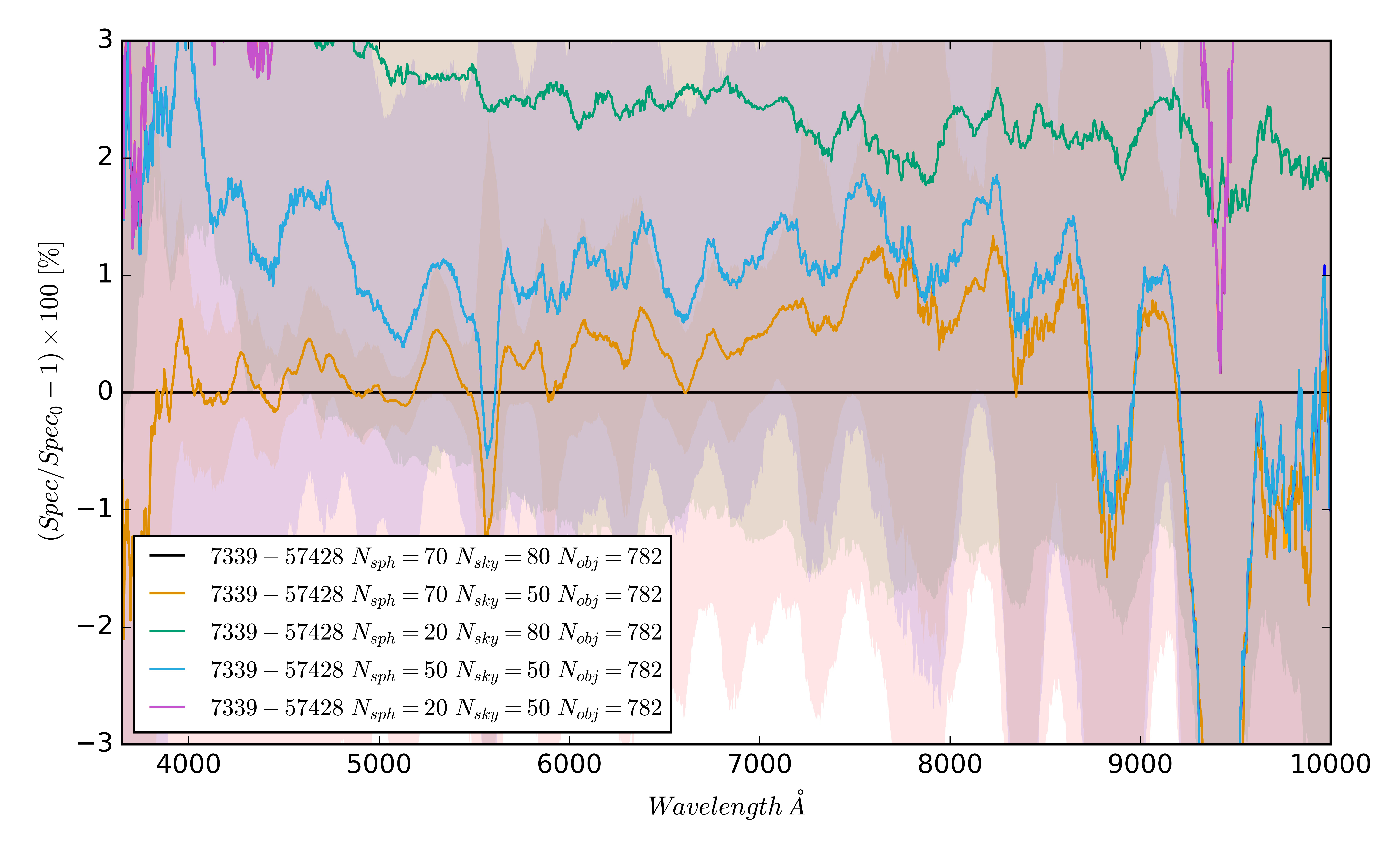}
	\caption{Comparison of the reduced spectrum of an object ($Spec$) when using the number of standards ($N_{sph}$) and skies ($N_{sky}$) in the legend relative to the spectrum of the object when using all of the calibrators ($Spec_0$). The sample shown is for plate-MJD 7339-57428. The left panel is for the 67 brightest galaxies ($m_r < 20$) on the plate and the right panel is the faintest galaxies ($m_r > 20$).}
	\label{fig:calib_number}
\end{figure*}

Figure \ref{fig:calib_number} shows the comparison of the reduced spectrum of an object ($Spec$) when using the number of standards ($N_{sph}$) and skies ($N_{sky}$) in the legend relative to the spectrum of the object when using all of the calibrators ($Spec_0$) for plate-MJD 7339-57428. The left panel is for the 67 brightest galaxies ($m_r < 20$) on the plate and the right panel is the faintest galaxies ($m_r > 20$). Overall, we can see that reducing the numbers of calibration fibers introduces an additional error of 1.5\% to 3\% for the low number of calibration fibers (20 standards and 50 skies for the two cameras) and 0.5\% to 1\% for the medium number of calibration fibers (50 standards and 50 skies for the two cameras).

Based on these results, we decided how many calibrators to use based on what was desired for a specific data collection scenario. For Bright Time and Dark Plane observations, we set the parameters to require the medium number of calibrators, as spectrophotometry is not as crucial for these results. Spectrophotometry is essential for Dark Monitoring and Dark RM, however, so the \texttt{designmode} was set to require a higher number of standards. The number of required sky fibers were not increased for these \texttt{designmode}'s, however, as Figure \ref{fig:calib_number} indicates that the number of standards generally has more of an influence on spectrophotometry than the number of skies. The Dark Faint \texttt{designmode}, which is optimized for observations of very faint galaxies, requires fewer standards but more sky fibers. The smaller number of standards is informed by the science goals of these fields, which do not require precise spectrophotometry.  The more stringent requirements on the number of sky fibers, however, are based on tests like that shown in  the right panel of Figure \ref{fig:calib_number}, where we can see that the addition of more skies results in a higher signal-to-noise spectrum than observations with the same number of standard stars but fewer sky fibers. Because of this, the number of skies is larger than normal to ensure adequate sky subtraction for the detection of these faint sources.

\subsubsection{Distribution of Calibrators in the Focal Plane}
\label{sec:calib_dist}
\begin{figure*}[!t]
	\centering
	\includegraphics[width=0.49\textwidth]{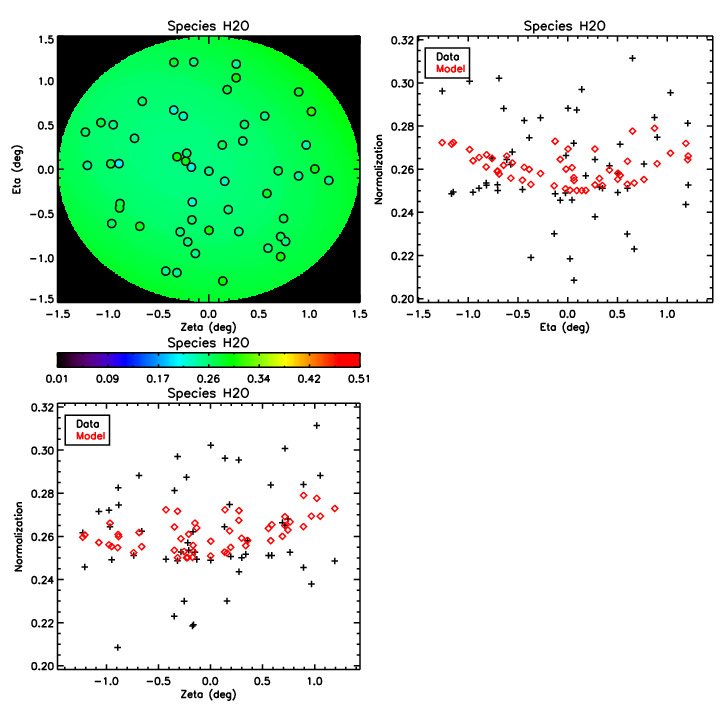}
    \includegraphics[width=0.49\textwidth]{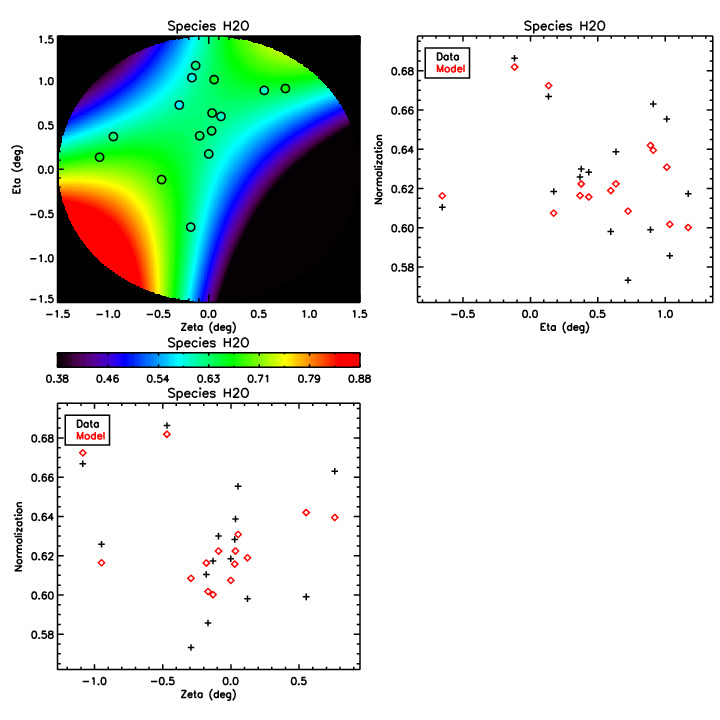}
	\caption{The model normalizing factor applied to account for telluric lines for two different scenarios. The left panel is for a larger number of APOGEE telluric standards well distributed across the FOV and the right panel is for a small number of standards not well distributed.}
	\label{fig:telluric_dist}
\end{figure*}

The FPS observes with a fairly large field of view (radius $\sim 1.5^\circ$ at APO, and $\sim 1^\circ$ at LCO). This means that the calibration (effective throughput, sky flux) can vary significantly for fibers on opposite sides of the FPS. Such changes need to be accounted for during reductions, which is accomplished by having standards and skies that are well distributed across the FOV. An example illustrating the importance of this consideration is shown in Figure \ref{fig:telluric_dist}. Here, the left panel shows the example of the modeled telluric correction for H$_2$O for an instance where there is a larger number of APOGEE telluric standards well distributed across the FOV. The right panel, in contrast, is for a smaller number of standards that are not well distributed across the field of view. It is clear that, especially when there are few standards in an observation, they must be well distributed across the field to fully constrain the fits that are used to interpolate each correction onto a given science fiber.

We tested metrics for the distribution of calibrators across the FPS, which we call the \texttt{skies\_FOV} and \texttt{stds\_FOV} \texttt{designmode} parameters, using a series of simulated \textit{Designs}. In each simulated \textit{Design}, we randomly assigned $N$ fibers for a given spectrograph to science and calibration targets, and computed the metric for that configuration. By doing this for many random realizations of these fiber selections, we could 1) develop the algorithmic form of our metric and 2) assess appropriate values to use in the metric parameters.

\begin{figure*}[!t]
	\centering
	\includegraphics[width=0.3\textwidth]{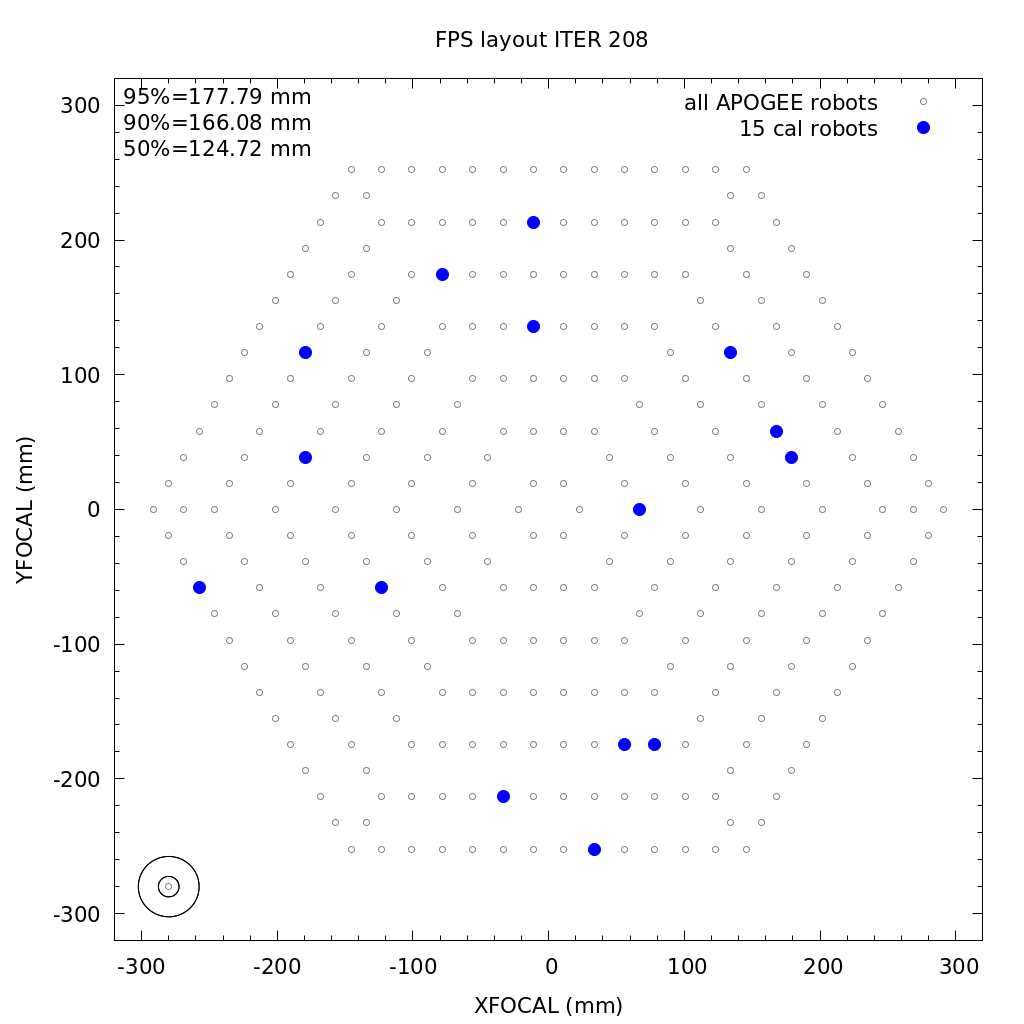}
    \includegraphics[width=0.3\textwidth]{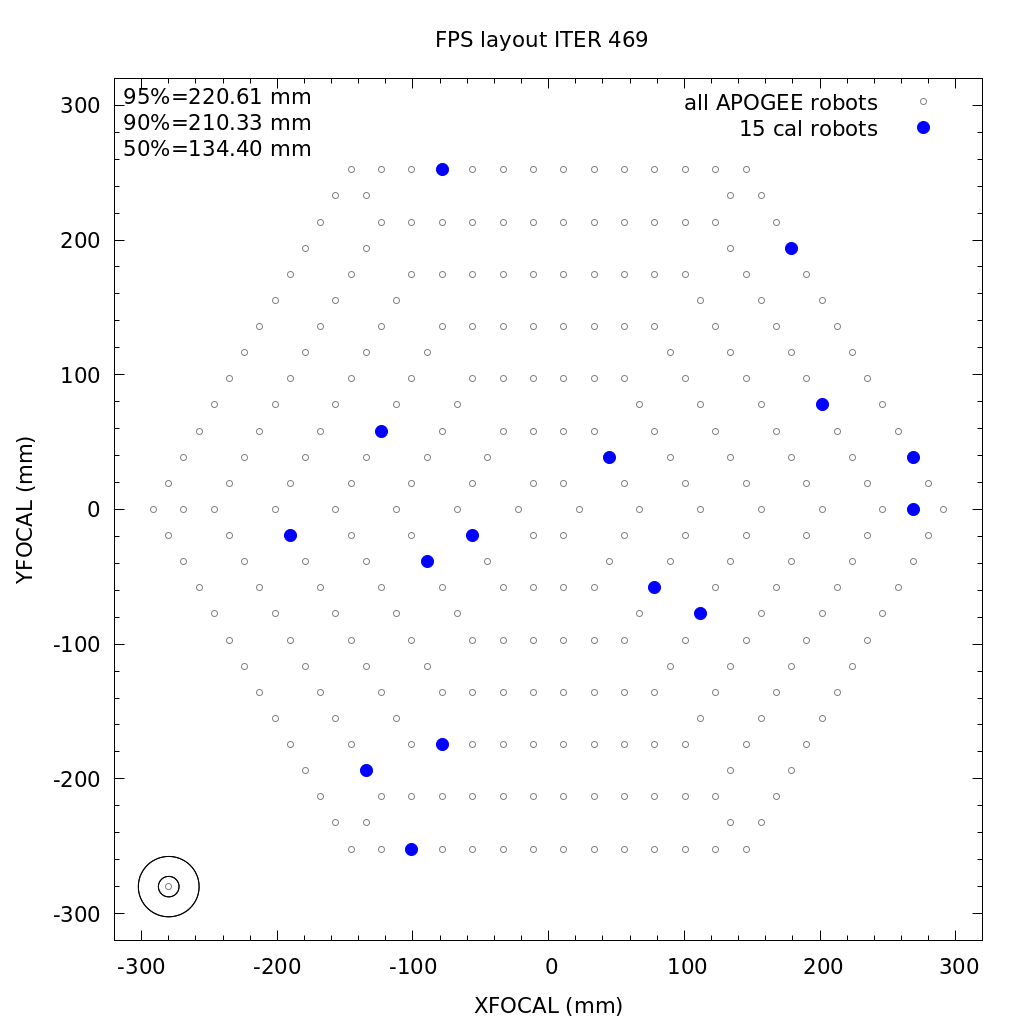}
    \includegraphics[width=0.3\textwidth]{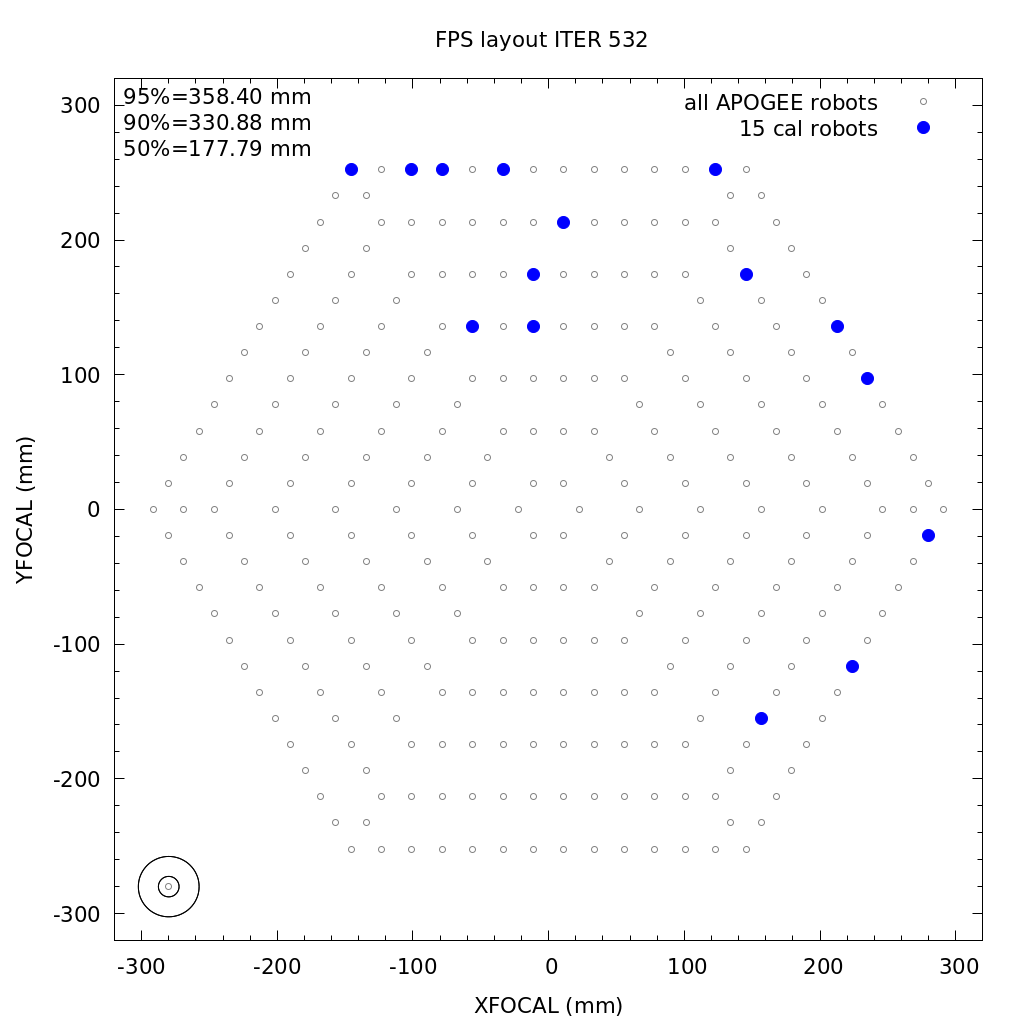}
	\caption{Examples of random selections of 15 APOGEE fibers in the FPS, where the 15 fibers selected represent fibers assigned to calibrators and are shown as the blue dots. The above examples show the best (left panel), typical (center panel) and worst (right panel) selection out of 1000 random selections when considering the 95th percentile of the distance from unassigned fibers to the 3rd closest assigned fiber.}
	\label{fig:fov_metric}
\end{figure*}

Figure \ref{fig:fov_metric} shows examples of random assignments of 15 APOGEE fibers in the FPS. Here these 15 fibers represent calibrator assignments for the \textit{Design} and all unassigned fibers would be non-calibrators. If we now consider the 95th percentile of the distance $d$ from unassigned fibers to the 3rd closest assigned fiber, which is the metric used for the \texttt{skies\_FOV} and \texttt{stds\_FOV} \texttt{designmode} parameters, the examples in Figure \ref{fig:fov_metric} show the best (left panel), typical (center panel) and worst (right panel) case scenarios. For the worst case scenario, it is clear that the calibrators are poorly distributed and heavily favor one side of the FPS, while the best and typical scenarios seem to have calibrators in all regions of the FPS.

\begin{figure}
    \centering
    \includegraphics[width=0.45\textwidth]{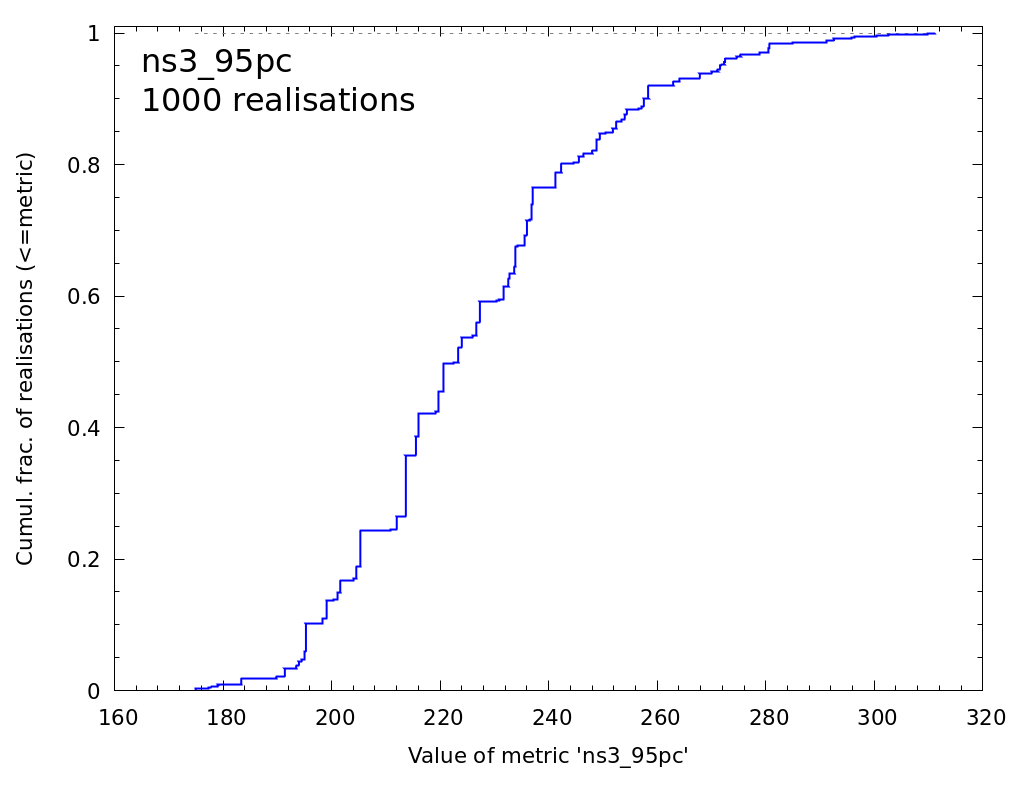}
    \caption{Cumulative distribution of the 95th percentile of the distance from unassigned fibers to the 3rd closest assigned fiber for 1000 random selections of 15 APOGEE fibers in the FPS. Examples of some of these random selections, and the values of their metrics, are also shown in Figure \ref{fig:fov_metric}.}
    \label{fig:fov_dist}
\end{figure}

Figure \ref{fig:fov_dist} shows the distribution of this metric for 1000 random selections of 15 APOGEE fibers in the FPS. If we were to rely on random assignments, it is clear that we would need to set the distance condition $d$ to a large value of $\sim300$ mm to pass $\sim99\%$ of \textit{Designs}. This would of course result in a large number of \textit{Designs} with non-uniform distributions across the field of view of the FPS, as is evident from the worst case scenario in Figure \ref{fig:fov_metric} (right panel). So, we instead usually chose to select a value for $d$ near the 50th percentile in these distributions, as we found this value resulted in an adequate distribution of standards. 
We reran these tests for each observing mode based on the minimum numbers of skies and standards required for each spectrograph, and evaluated the median of similar cumulative distributions as in Figure \ref{fig:fov_dist}. We construct a metric where the distance id calculated to the $k=1$ neighbor for \texttt{designmode} parameters that had a larger number of minimum skies or standards, and $k=3$ for ones with smaller minima.

\begin{figure}[!t]
	\centering
	\includegraphics[width=0.45\textwidth]{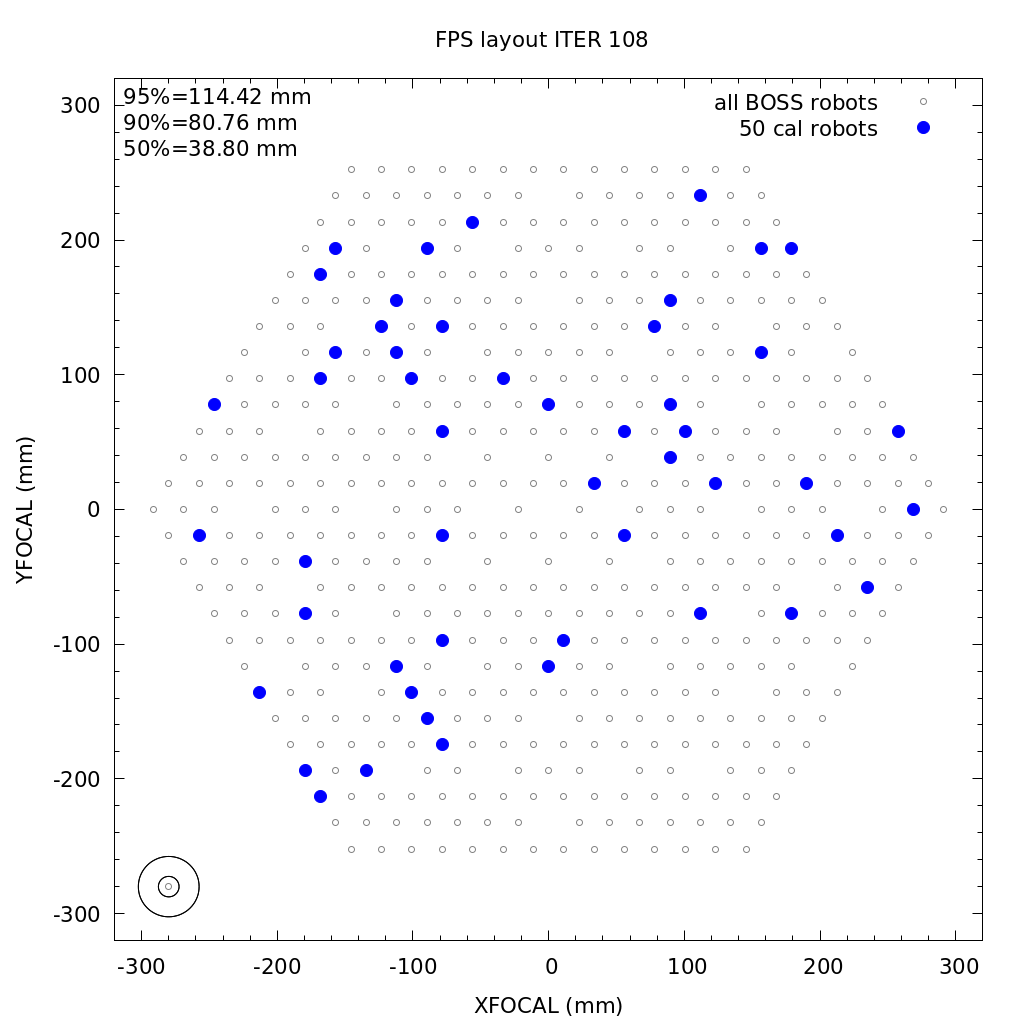}
	\caption{Worst distributed example of random selections of 50 BOSS fibers in the FPS, where the 50 fibers selected represent fibers assigned to calibrators and are shown as the blue dots. The above examples show the worst selection out of 1000 random selections when considering the 95th percentile of the distance from unassigned fibers to the 1st closest assigned fiber.}
	\label{fig:fov_metric_boss}
\end{figure}

While we were able to set values for this metric for each \texttt{designmode}, we deemed it was not always necessary to constrain the assignments in the \textit{Design} based on this distribution metric. For example, Figure \ref{fig:fov_metric_boss} shows the simulated scenario with the worst spatial distribution (out of 1000 realizations), for 50 BOSS fibers randomly assigned in the FPS. Here, the metric considered is the 95th percentile of the distance from unassigned fibers to the 1st closest assigned fiber, similar to what is used for most BOSS related FOV metrics in Table \ref{tab:targetdb-designmode-values}. Even in this worst simulated scenario, the BOSS calibrators are fairly well distributed across the field of view with assignments only missing in a small area of the FPS. Because of this, we did not deem it necessary to apply the BOSS related FOV constraints to limit real survey \textit{Designs}. 

While we are not currently enforcing requirements on this metric on all \textit{Designs}, we do compute and track the metric throughout the many iterations of survey planning. So, even though this metric is not currently used to constrain the BOSS standards within \textit{Designs}, it is still an interesting parameter to monitor and ensure it does not, on average, become too high as we change the assignment logic within the survey planning software. Additionally, we have found throughout the course of the survey that the distribution of BOSS standards is not as strongly correlated with the quality of the spectrophotometry as we had initially expected.

\begin{figure*}[!t]
	\centering
	\includegraphics[width=0.95\textwidth]{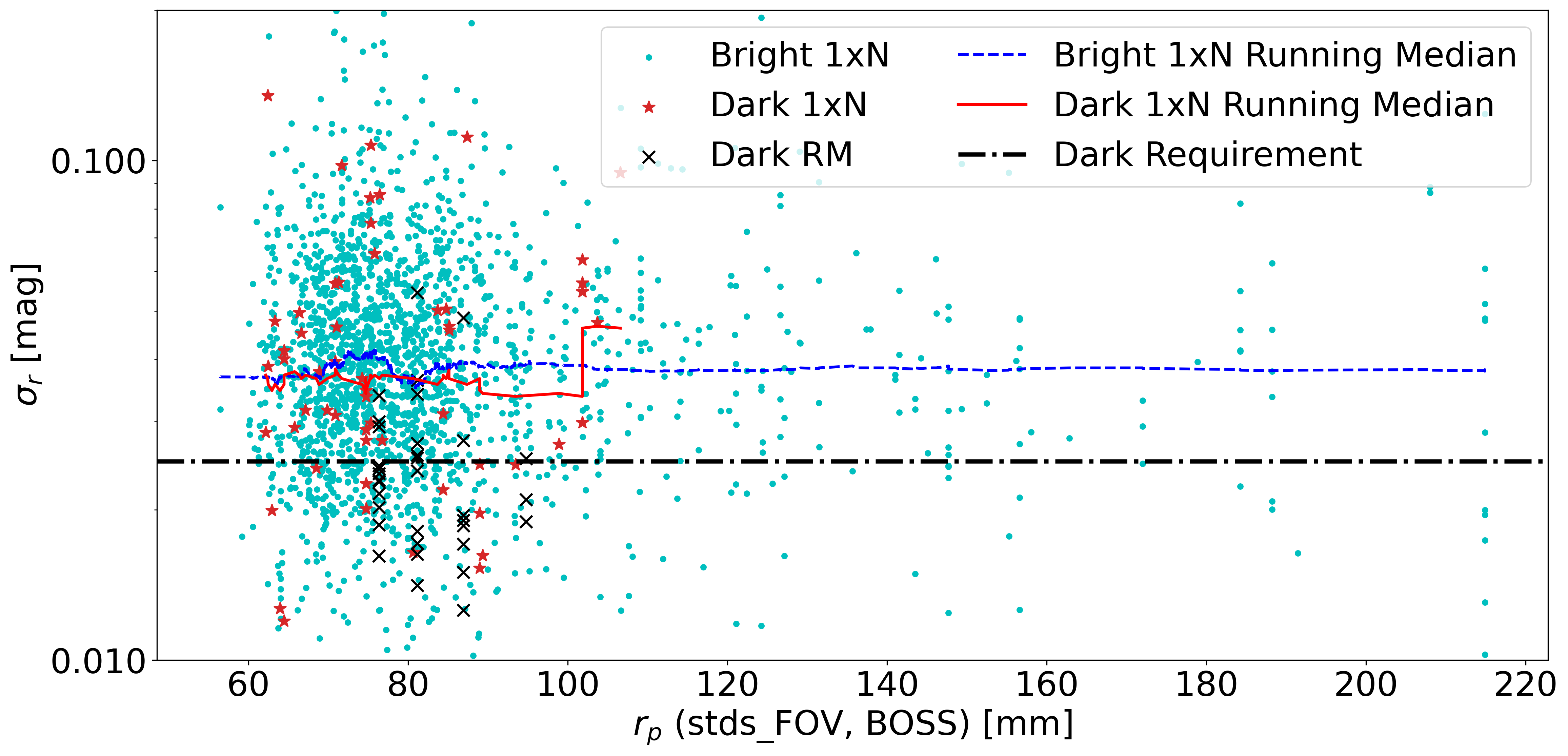}
	\caption{The standard deviation in the r-band $|$\texttt{CALIBFLUX}--\texttt{SPECTROFLUX}$|$ vs.~$r_p$ as calculated for the \texttt{stds\_FOV} metric for BOSS \textit{Designs} in Bright Time (green), Dark Plane (red stars) or Dark RM (blue $\times$) observed at APO and LCO for MJD$ > 60654$. Here, \texttt{CALIBFLUX} is the predicted flux in nMgy down the fiber of the science or calibration target based on SDSS photometry and \texttt{SPECTROFLUX} is the flux derived form the 1D spectra. The green and red lines show the running median for the Bright Time and Dark Plane \textit{Designs}, respectively. The black line shows the spectrophotometric requirement for dark time observations, where \textit{Designs} need to be less than this value to meet the criteria.}
	\label{fig:boss_fov_observed}
\end{figure*}

Figure \ref{fig:boss_fov_observed} shows the standard deviation in the r-band $|$\texttt{CALIBFLUX}--\texttt{SPECTROFLUX}$|$ vs.~$r_p$ as calculated for the \texttt{stds\_FOV} metric for BOSS \textit{Designs} observed for MJD$ > 60654$. Here, $\sigma_r$ is used to assess the quality of the spectrophotometry during dark time scenarios, where the requirement is shown as the black line in Figure \ref{fig:boss_fov_observed}. Overall, we see that the distributions of standards have little effect on this spectrophotometry metric. This is reinforced by the Dark RM \textit{Designs} shown in Figure \ref{fig:boss_fov_observed}. Here, each vertical row of X's is one Dark RM \textit{Field} that contains repeated observations of the same \textit{Design}. So, even with the same \textit{Design} (and thus same number and distribution of standards), we see a large variance in the spectrophotometry. This means that other factors, such as observing conditions and accuracy of fiber positions, have a much larger impact on the spectrophotometry than this one constraint.

From Figure \ref{fig:fov_metric}, it is clear that the distribution of standards across the FOV is much more of a concern for APOGEE reductions, however, as there are usually very few standards assigned per \textit{Design}. As a result, we must impose some constraint on the APOGEE standard assignments to ensure a reasonably uniform distribution across the FPS. In the following section, we will outline how we implement this constraint to get favorable results during survey planning.

\subsubsection{Implementing FOV Constraint in Practice}

As it can be seen in the previous section, calibrators can be assigned randomly and result in a uniform distribution across the FPS according to our metric for the \texttt{designmode}. It is of course inefficient to randomly create configurations of fibers until this criteria is reached, especially when there are a number of other constraints on the \textit{Design}. This is why for the APOGEE standards, the one calibrator we are most concerned about being well distributed when fibers are assigned randomly, we crafted a different assignment logic.

\begin{figure*}[!t]
	\centering
	\includegraphics[width=\textwidth]{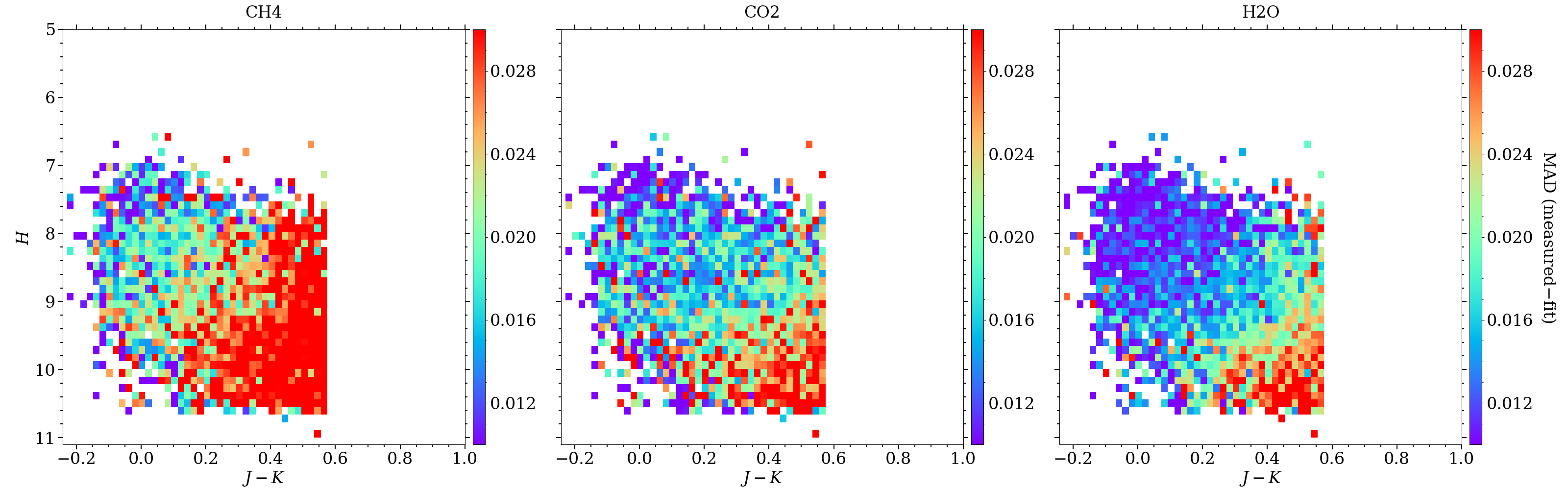}
	\caption{Difference between telluric features determined for individual stars in a \textit{Field} and the smooth spatial fit determined from all stars in the \textit{Field}, as a function of brightness and color of the individual stars. Worse fits are observed for fainter and redder sources.}
	\label{fig:apogee_tell_colors}
\end{figure*}

In this assignment logic, we not only have a preferred distribution of APOGEE standards across the FPS, we also have a preferred distribution of APOGEE standard colors and magnitudes. To accomplish the desired distribution in color and magnitude, the standards are sorted by their ``goodness", where:
\begin{equation}
\mathrm{goodness} = - 13.33\left[(J-K) - 0.25\right] - (H - 9),
\end{equation}
which prefers brighter and bluer stars. This condition was set based on experiments comparing how well telluric features determined for individual stars in a \textit{Field} agreed with the smooth spatial fit determined for all stars in the \textit{Field}, as a function of brightness and color of the individual stars. This is summarized in Figure \ref{fig:apogee_tell_colors}. The residuals measured in these experiments showed a trend in brightness and color roughly corresponding to the above equation.

\begin{figure}[!t]
	\centering
	\includegraphics[width=0.45\textwidth]{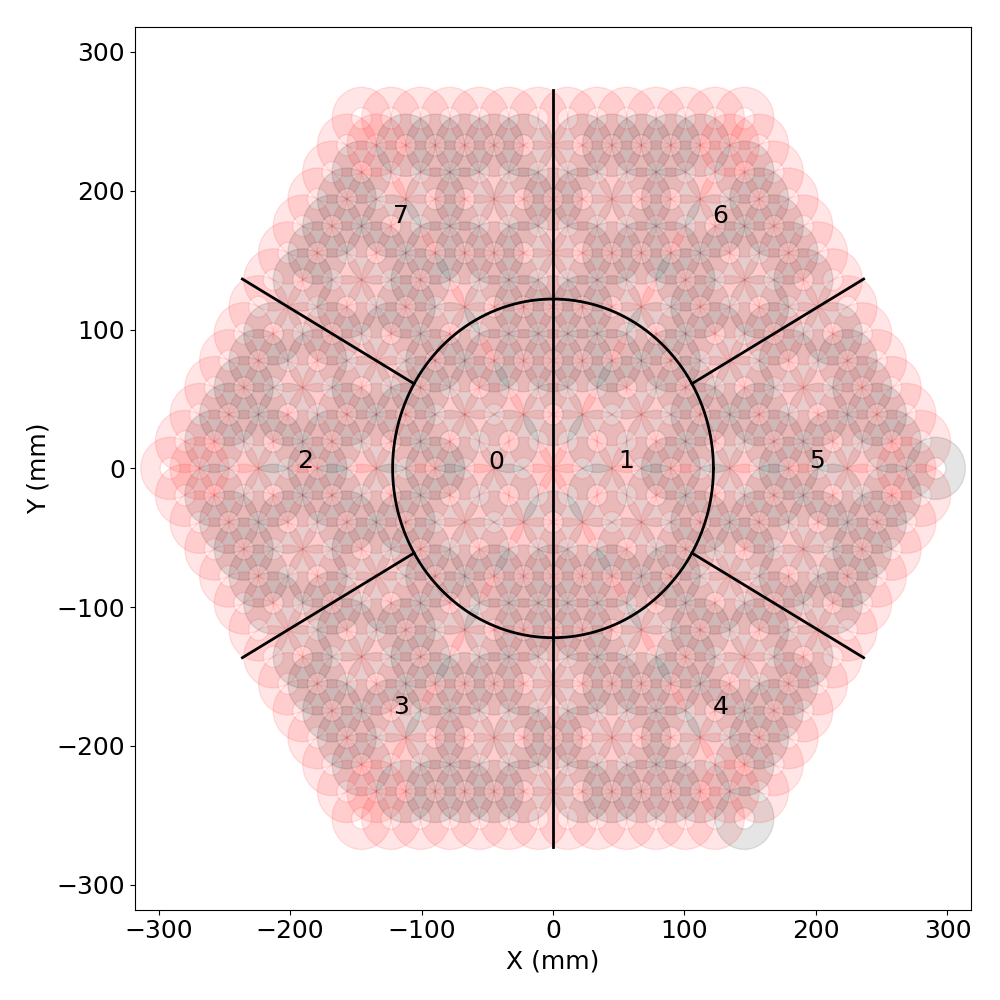}
	\caption{Layout of the focal plane for the FPS system
  used by SDSS-V. The layout shown is as-built for APO; the LCO layout
  is nearly identical.  The $X$ and $Y$ axes are position in the focal
  plane, with the boresight at $X=Y=0$ mm. We show each positioner as
  an annulus describing its patrol area. The pink annuli are the 298
  positioners that carry both BOSS and APOGEE fibers. The grey annuli
  are the 202 positioners that carry a BOSS fiber but not an APOGEE
  fiber.  The focal plane is divided into eight zones, as labeled, for
  the purposes of distributing APOGEE standard stars.}
	\label{fig:apogee_zones}
\end{figure}

To add the additional constraint of FPS location on these standards, we construct 8 ``zones" in the FPS. This is a similar approach as from the past APOGEE surveys \citep{Zasowski2013, Zasowski2017}. These zones are shown in Figure \ref{fig:apogee_zones} as the solid black lines and each red annulus is the patrol radius for an APOGEE fiber in the FPS. Using these zones, at the beginning of the fiber assignment for any \textit{Field} we determine the goodness threshold to use for selecting APOGEE standard in each zone in Figure \ref{fig:apogee_zones}. We use a goodness threshold of zero by default, but if there are less than 3 stars available above the threshold, we reduce the threshold until there are enough standards available. The assignment process (which requires assigning at least one APOGEE standard to each zone, and 15 APOGEE standards total) then has enough to choose from among the best available APOGEE standards.

\begin{figure*}[!t]
	\centering
	\includegraphics[width=0.9\textwidth]{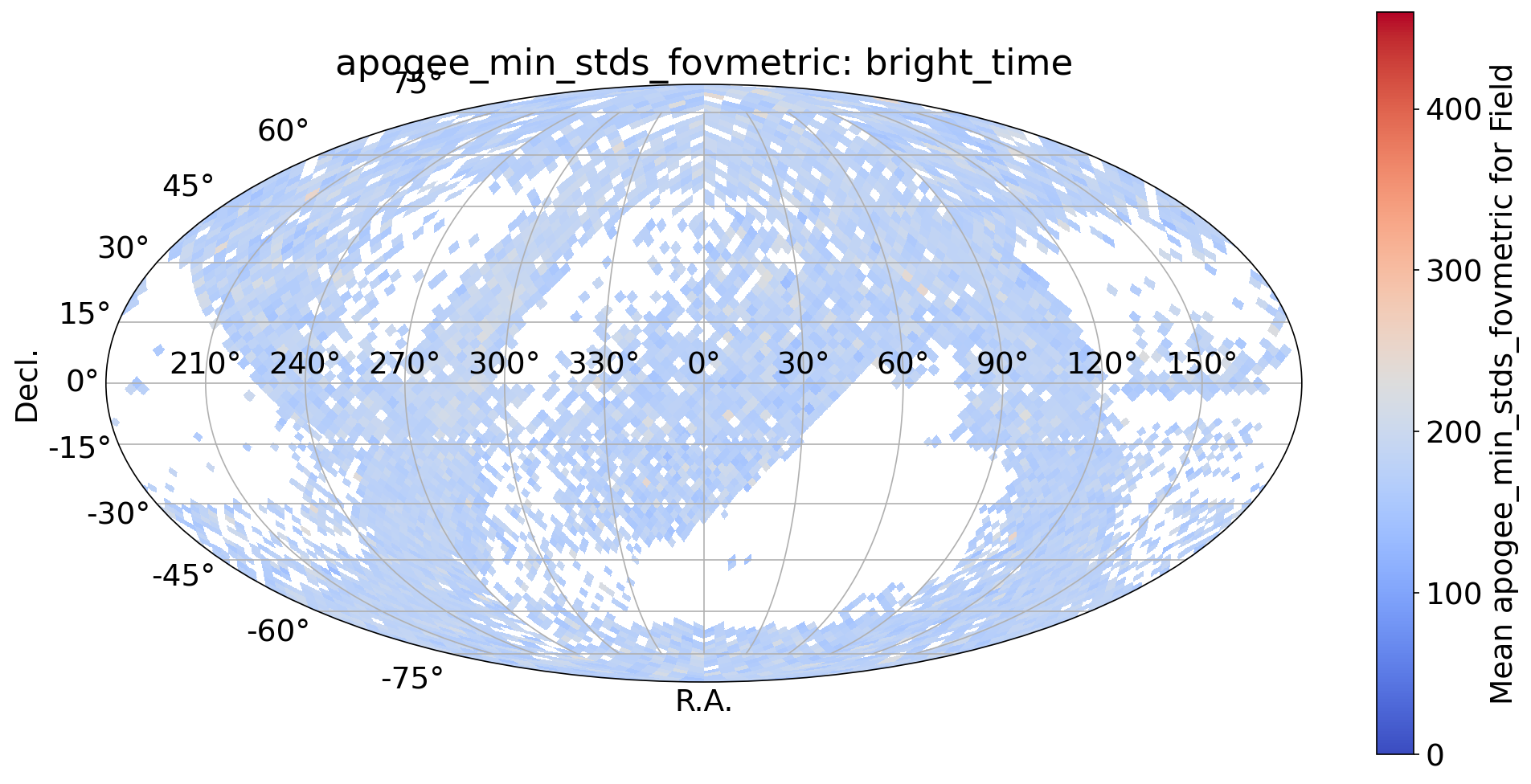}
    \includegraphics[width=0.9\textwidth]{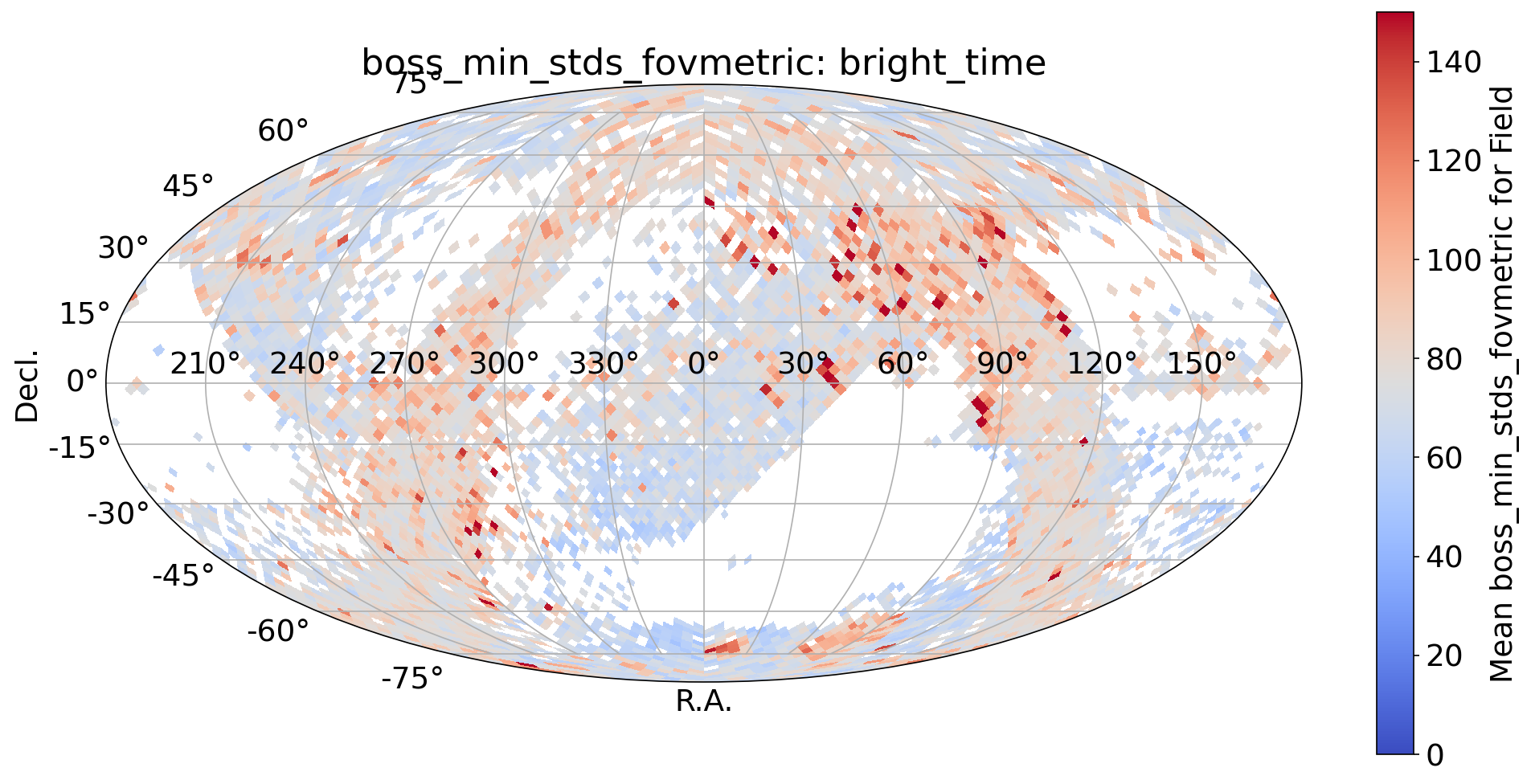}
	\caption{The mean value of \texttt{stds\_FOV} for APOGEE (top panel) and \texttt{stds\_FOV} for BOSS (bottom panel) in Bright Time for the current survey plan for SDSS-V. In both plots, \textit{Fields} colored in blue mean on average the \textit{Designs} pass our FOV metric. Overall, $\sim97\%$ of APO \textit{Designs} and $\sim99\%$ of LCO \textit{Designs} pass our FOV metric criteria when assigning the APOGEE standards using the zones in Figure \ref{fig:apogee_zones}. A much lower rate of $\sim42\%$ for APO and $\sim46\%$ for LCO pass the FOV metric for BOSS standards where such zoning is not required during the assignment stage.}
	\label{fig:apogee_fov_valid}
\end{figure*}

To test the validity of this assignment logic, when creating a survey plan we always check the FOV metric for the \textit{Designs}. The top panel of Figure \ref{fig:apogee_fov_valid} shows the mean value of \texttt{stds\_FOV} for APOGEE in Bright Time for the current survey plan for SDSS-V. Here, the colorbar is scaled such that \textit{Fields} colored in blue indicate on average the \textit{Designs} pass our FOV metric. Overall, $\sim97\%$ of APO \textit{Designs} and $\sim99\%$ of LCO \textit{Designs} pass our FOV metric criteria when assigning the APOGEE standards using the zones in Figure \ref{fig:apogee_zones} for this survey plan. Similar results are found for other dark time observing modes as well.

We do note that for the BOSS standards, where we do not implement this zone-based criterion for assignments, pass rates can be much lower. However, this seems to only occur for \textit{Fields} in the Galactic plane (bottom panel of Figure \ref{fig:apogee_fov_valid}). This is evident considering the Dark Plane observing mode, which is heavily concentrated in the Galactic plane, and for which only $\sim50\%$ of \textit{Designs} pass the FOV metric for BOSS standards. For observing modes like Dark Monitoring and Dark Faint, which are primarily at high Galactic latitudes, pass rates with this random, unconstrained assignment are $>90\%$ and $>70\%$, respectively. To reiterate, the number of BOSS standards requested is typically large, so even with this poor metric results, distributions are adequate for reduction purposes. For example, $\sim90\%$ of \textit{Designs} have a FOV metric better than the worst case scenario in Figure \ref{fig:fov_metric_boss} for Bright Time in these survey simulations. Finally, we again want to emphasize that for BOSS it seems that this metric is not strongly correlated with the quality of the spectrophotometry (Figure \ref{fig:boss_fov_observed}).

\subsubsection{Magnitude Limits for Assignments}\label{sec:mag_limits}


There are two main considerations when determining the magnitude limits for a \textit{Design}; at what magnitude \textit{all} reductions become impossible (due to saturation of standards either directly or indirectly from contamination) and at what magnitude \textit{some} reductions become impossible (e.g., at the faint end due to contamination of light from neighboring bright stars on the chip). For the former, this is the main consideration during Bright Time because we do not plan to observe stars at the very faint magnitude end of the dynamic range of the survey. For the dark time observing modes, the latter consideration is crucial because it is in these dark sky conditions we typically observe the faintest objects, as well as the objects requiring the best spectrophotometric accuracy. So, here we will consider these two classes of observing modes separately.

Generally, most stars targeted by MWM with BOSS during Bright Time have $G > 12.5$. Previous SDSS programs, like MaNGA have set the bright limit for target selection to 12.7 mag in the $g$ and $i$ bands \citep{yan2019}. This was because at this magnitude, the maximum count in the blue or red camera would be $\sim$25,000 based on the throughput of the instrument \citep{yan2016}, and the detector begins to exhibit nonlinearity above 33,000 counts. So, this limit conservatively avoids this nonlinear response. Similar limits were placed on targets during the SDSS-V plate observations. With these limits, we found that during plate observations a linear response between the expected and measured flux was maintained. As this is the main concern for targets observed in Bright Time, this same bright limit was set for the relevant Bright Time \texttt{designmode} parameters. This does limit the observation of targets brighter than this limit.
As will be discussed later in the paper (see Section \ref{sec:offsets}), we will develop another mode of observation that will faciliate the observations of brighter targets, while still allowing for proper reductions.

For BOSS targets in dark time, the setting of a bright limit is crucial for more than just ensuring that the response on the chip is in the linear regime. Indeed, for many of these data collection scenarios we will be targeting very faint sources, so flux from adjacent traces on the chip can have a great influence on the reductions for these faint objects. The BOSS pipeline does attempt to model the contamination from neighboring fibers, however i) this modeling is not perfect, and ii) contaminating flux from neighbors will increase noise in extracted spectra, even when modeling is perfect.
To assess this, we used SDSS-V plate data to estimate the level of contamination of extracted spectra associated with sky fibers by bright neighboring targets. Here we selected plates with \texttt{PLATESN2} $>10$, which resulted in 9814 BOSS sky fibers from SDSS-V plates. For these sky fibers, we found all nearest neighbors one trace up and down on the chip. We then examined the observed residual flux of the sky spectrum as a function of the magnitude of the brightest neighbor. In the absence of contamination, and for a perfect sky subtraction algorithm, the observed flux should be around zero when averaged over many pixels. Any flux greater than this should then be a good estimator of the residual contamination from the on chip neighbor.

\begin{figure*}[!t]
	\centering
	\includegraphics[width=0.45\textwidth]{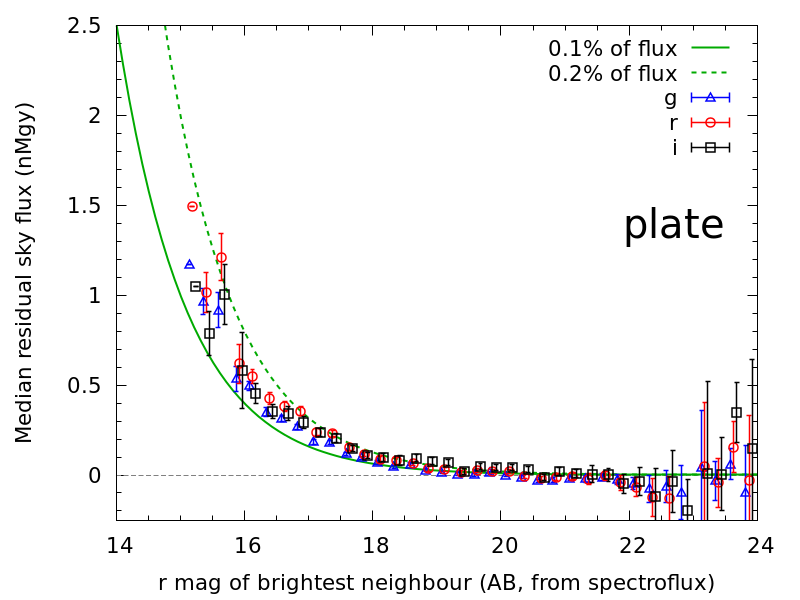}
	\includegraphics[width=0.45\textwidth]{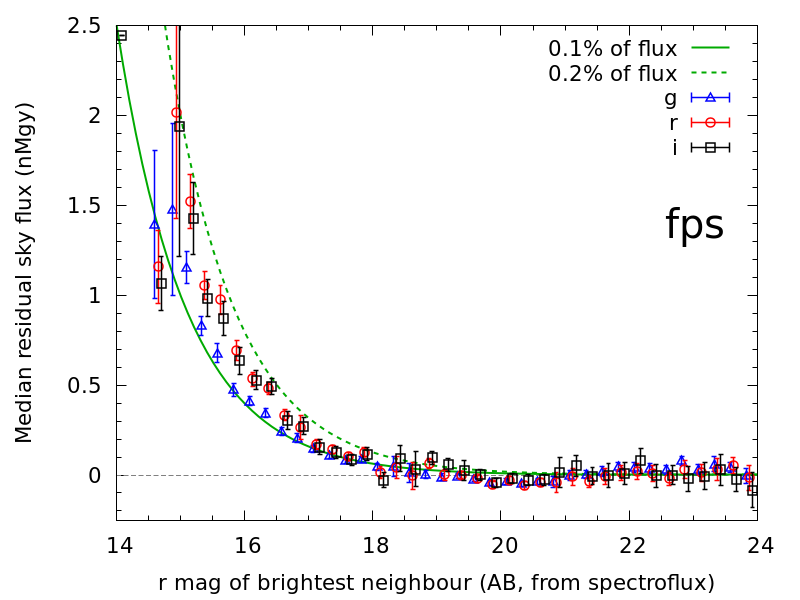}
	\caption{Median residual spectral flux for sky fibers (averaged over the SDSS $g$, $r$ and $i$ bandpasses), as a function of the magnitude of the brightest on chip neighbor (estimated from the observed spectroflux). The left panel shows the trend for DR19 \citep{sdssdr19} dark-time plate data, and the right panel for the DR19 FPS dark-time dataset. The blue, red and black data points are the median of the bandpass-averaged sky fiber flux for the $g$, $r$ and $i$ bands respectively. Here, the sky spectra have been sky subtracted, so any flux greater than zero in the sky spectrum probes the contamination from the on chip neighbor (N.B. in order to correct for a known small systematic over-subtraction of sky flux in DR19 BOSS data, a zeropoint correction of +0.18\,nMgy has been applied to all y-axis values). The green lines show the residuals expected if 0.1 (solid) or 0.2\% (dashed) of the bright neighbor flux were to remain un-subtracted from the sky fiber spectra.}
	\label{fig:cross_talk_bright_limit}
\end{figure*}

Figure \ref{fig:cross_talk_bright_limit} shows the average residual flux of sky fibers (averaged over the bandpasses of the SDSS $g$, $r$ and $i$ filters) as a function of the magnitude of the brightest on chip neighbor. The faintest dark-time science targets in SDSS-V reach to around $1-2$ nMgy ($\sim22$ AB). From Figure \ref{fig:cross_talk_bright_limit}, we see that the residual contaminating flux is brighter than 1\,nMgy for neighboring targets brighter than $\sim16$ mag. Because of this, for most dark time data collection scenarios the bright limit will be 16 mag to ensure that \edit1{they} do not adversely impact the spectral quality for our faintest targets. The exception is in the Dark Plane observing mode, because objects targeted in those \textit{Designs} are not as faint as the faintest BHM targets. Figure \ref{fig:cross_talk_bright_limit} also illustrates that the relative strength of on-chip crosstalk contamination has not changed markedly with the upgrade from plug plates to the FPS, as expected.


For both of the above scenarios with BOSS, the science and standard targets have similar minimum magnitude thresholds for the same reasoning outlined above. We do not set a maximum magnitude for the science targets, as we assume the individual programs will appropriately determine their faint limits based on their individual science goals. On the other hand, the standards have to be higher signal-to-noise ratio to be useful for reductions. Because of this, we do set a maximum allowable magnitude for these. 
We impose a faint magnitude limit of $r_\mathrm{psf} < 18$\,AB for BOSS standard stars. This limit is a compromise between i) the need to obtain a sufficient $S/N$ per star within a typical science exposure, and ii) the desire to minimize the impact of residual on-chip contamination from neighboring spectra, both balanced against the challenge of finding enough suitable spectrophotometric standard stars per \textit{Field} at high Galactic latitudes. 

For all data collection scenarios with APOGEE, we relied on past experience with the instrument during SDSS-IV. It was found during the commissioning for APOGEE during SDSS-IV that due to saturation limits with detectors, combined with unexpected superpersistence problem on regions of some of the detector arrays, that a bright limit of $H > 7$ was advisable \citep{Zasowski2013}. We use the same bright limit here for all data collection scenarios for both standards and science targets. Specifically for the standards, we need enough signal for the reductions to be successful, so a maximum (faint) limit is set at $H=13$ for all data collection scenarios for the standards.

\section{Special Considerations for Bright Stars}\label{sec:bright}

\subsection{Fiber Offsets to Intentionally Target a Bright Star}\label{sec:offsets}
\label{sec:fiber_offsets}
For the various data collection scenarios, bright limits are set on observations to avoid saturating the CCD in extreme cases, but more importantly to avoid cross-contamination of light on the chip (see Section \ref{sec:mag_limits}). However, without a way to target objects having brighter magnitudes, this choice would limit what can be observed for SDSS-V. In order to observe these brighter objects, one solution is to deliberately offset the fiber from the target such that the flux entering the fiber is below the bright limit for the observation. To do this accurately, the PSF of the FPS has to be well-known. In the sections below, we discuss the tests performed to describe the PSF and how we use this to implement offsets for SDSS-V. 

\subsubsection{Determining PSF Wings}\label{sec:wings}


First we focus on the PSF ``wings", i.e.~its values at larger angular distances.
A PSF can extend over large angular distances \citep[up to 8 arcmin in the SDSS photometric survey;][]{psfwings} and must be characterized to inform bright star avoidance rules (discussed in Section \ref{sec:bright_avoid}). In subsequent sections we will discuss the modeling of the PSF ``core", i.e.~its values at smaller angular distances, which are dominated by the instantaneous properties of the atmosphere and will be crucial for deliberate offsetting from brighter objects.

To examine the shape of the PSF wings, we use archival data from the BOSS and eBOSS surveys \citep{SDSS_DR16}. These surveys were conducted with a different system than SDSS-V; they only used the Sloan Foundation Telescope at APO, and at that time the telescope used a different spectroscopic corrector and used the plug-plate system. Nevertheless, we expect that this analysis yields a good approximation for the shape of the PSF wings for SDSS-V.

For this analysis, we select all eBOSS/BOSS optical spectra that happen to lie within 90" of a bright ($V_T < 12$ mag) Tycho-2 star. In order to determine the excess flux down the fibers to constrain the PSF wings of these bright Tycho-2 sources, we need to compare the \texttt{CALIBFLUX} (predicted flux in nMgy down the fiber of the science or calibration target based on SDSS photometry) to the \texttt{SPECTROFLUX} (flux derived form the 1D spectra). Any flux difference between these two values should then give the excess flux that can be attributed to the bright Tycho-2 source.

To verify that \texttt{CALIBFLUX} is a good predictor of \texttt{SPECTROFLUX} in the absence of a bright source, we select all eBOSS/BOSS spectra that are $>90$" from the a Tycho-2 star. This results in 3,750,720 eBOSS/BOSS entries. 
From this sample we find that \texttt{CALIBFLUX} is overestimating \texttt{SPECTROFLUX} by $2-4\%$ with a weak dependence on \texttt{CALIBFLUX}. This would result in an artificial suppression of the PSF wings in our analysis. If we only consider stars with $\texttt{CALIBFLUX} < 2.5$ nMgy though, this effect seems to be much less and the difference between the measurements is at the $0-1\%$ level in the r- and i-bands. For the subsequent analysis then, we will only consider eBOSS/BOSS spectra that happens to be within 90" of a bright ($V_T < 12$ mag) Tycho-2 stars where $\texttt{CALIBFLUX} < 2.5$ nMgy (i.e. $>21.5$ mag), which results in 57,183 spectra.

\begin{figure}[!t]
	\centering
	\includegraphics[width=0.49\textwidth]{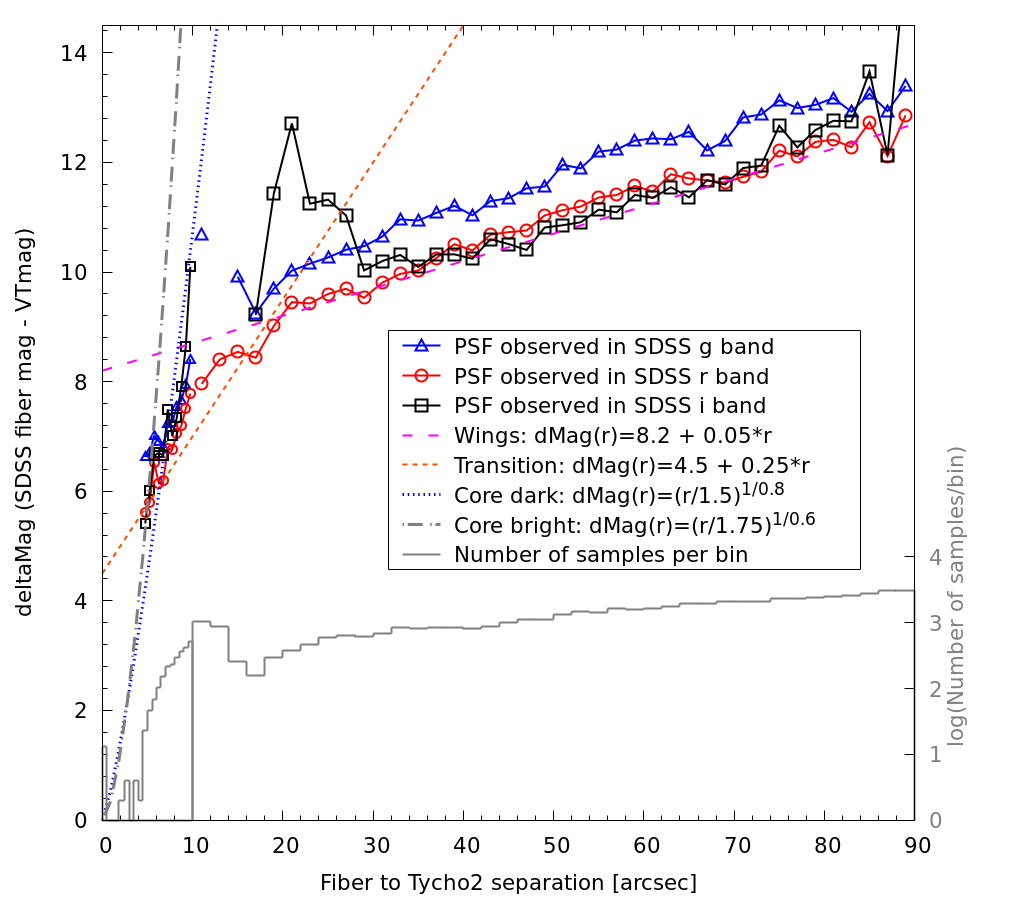}
	\caption{Median difference between excess magnitude of eBOSS/BOSS spectra (\edit1{derived from} $\texttt{SPECTROFLUX}-\texttt{CALIBFLUX}$) and the magnitude of the nearby Tycho-2 source, versus angular distance to the Tycho-2 source. For separations $<10$" no quality cuts are applied due to low numbers in the sample, but for separations $>10$" only spectra with $\texttt{CALIBFLUX} < 2.5$ nMgy are considered. The \edit1{points} show the median differences in the g-band (\edit1{blue triangles}), r-band (red \edit1{circles}) and i-band (black \edit1{boxes}). The histogram shows the number of spectra per bin of \edit1{fiber-to-star separation}. The dashed lines show the empirical fits to the median differences \edit1{for the r-band dataset}.}
	\label{fig:psf_wings}
\end{figure}

Using this quality cut, Figure \ref{fig:psf_wings} shows the median difference between excess magnitude of eBOSS/BOSS spectra (defined as $\texttt{SPECTROFLUX}-\texttt{CALIBFLUX}$) and the magnitude of the nearby Tycho-2 source versus angular distance to the Tycho-2 source. We do note that for separations $<10$" no quality cuts are applied due to low numbers in the sample, but for separations $>10$" only spectra with $\texttt{CALIBFLUX} < 2.5$ nMgy are considered. Additionally, we found in our analysis that there may be some correlations between this relative flux excess and azimuth around the object, but we are ignoring such structure here. In all bandpasses in Figure \ref{fig:psf_wings}, we  observe three apparent regions in the shape of the PSF; the ``core" ($\theta \lessapprox 6$"), the ``transition" ($6 \lessapprox \theta \lessapprox 20$") and the ``wing" ($\gtrapprox 20$"). These various relationships seem the most apparent in the r-band. With these r-band median differences, we fit functional forms that seemed to closely bound the data. In the core, we defined two relationships simply so the bright time function was less conservative than the \edit1{dark} time one. The resulting functions for each region are:
\begin{equation}\label{eq:core_bright}
    \Delta {\rm mag}_{\rm core, bright} = \left(\frac{r}{1.75}\right)^{1/0.6}
\end{equation}
\begin{equation}\label{eq:core_dark}
    \Delta {\rm mag}_{\rm core, dark} = \left(\frac{r}{1.5}\right)^{1/0.8}
\end{equation}
For the transition and wings area, we use a linear fit with no difference between bright and dark time:
\begin{equation}\label{eq:trans}
    \Delta {\rm mag}_{\rm transition} = 4.5 + 0.25 \times r
\end{equation}
\begin{equation}\label{eq:wing}
    \Delta {\rm mag}_{\rm wing} = 8.2 + 0.05 \times r 
\end{equation}
In the above equations, $r$ is the separation in arcseconds.

The shape of the PSF is described as a piece-wise function where the expected flux for some separation is the minimum value from the $\Delta$mag relations in eq. \ref{eq:core_bright}$-$\ref{eq:wing}. For the initial survey plans, these equations were used to describe the PSF for the implementation of offsets and  bright star avoidance. Due to the low sampling in the core region, however, we did not consider the relations in Eq. \ref{eq:core_bright} and \ref{eq:core_dark} a sufficiently good representation of the core PSF. In subsequent tests described below, we will arrive at better fits for these regions and these formulations of the core will be used in future survey plans. A summary of which descriptions of the PSF were used for various iterations of the survey plan is in Table \ref{tab:offset_versions}.  Overall, we consider the results in this section a good approximation for the wings of the PSF, and that they will be sufficient in describing the PSF at these large separations for the entirety of the survey.

\begin{table*}
  \footnotesize
  \centering
  \caption{Which equations were used to describe the PSF for offsetting and/or bright star avoidance during the various version of the survey plan. The final column indicates if offsetting was allowed during that version of the survey plan for each lunation type. A ``\checkmark" indicates it was allowed and an ``\textbf{X}" indicates it was not. In all cases, the bright star avoidance was implemented using the set of equations for a given version.} 
  \begin{tabular}{|l|c|c|c|c|}
      \hline
       \textbf{Survey Plan Version} & \textbf{Core} & \textbf{Transition} & \textbf{Wings} & \textbf{Offsetting} \\
      \hline
      \hline 
      \texttt{zeta} series (DR19) & Eqs.~\ref{eq:core_bright} \& \ref{eq:core_dark} & Eq.~\ref{eq:trans} & Eq.~\ref{eq:wing} & \makecell{Bright \texttt{designmode}'s: \textbf{X} \\ Dark \texttt{designmode}'s: \textbf{X}} \\
      \hline 
      \texttt{eta} series (DR20) & \makecell{Moffat FWHM$=1.7''$ and $\beta = 5$ \\ \quad at APO,\\Moffat FWHM$=1''$ and $\beta = 2$ \\ \quad at LCO} & Eq.~\ref{eq:trans} & Eq.~\ref{eq:wing} & \makecell{Bright \texttt{designmode}'s: \checkmark \\ Dark \texttt{designmode}'s: \textbf{X}} \\
      \hline 
      \texttt{theta} series (DR20 \& DR21) & \makecell{Moffat FWHM$=0.5''$ and $\beta = 1.6$ \\ \quad at APO (Bright \texttt{designmode}'s),\\Moffat FWHM$=1.4''$ and $\beta = 1.9$ \\ \quad  at APO (Dark \texttt{designmode}'s),\\Moffat FWHM$=0.8''$ and $\beta = 1.7$ \\ \quad  at LCO (Bright \texttt{designmode}'s),\\Moffat FWHM$=1.1''$ and $\beta = 1.8$ \\ \quad  at LCO (Dark \texttt{designmode}'s)} & \nodata & Eq.~\ref{eq:wing} & \makecell{Bright \texttt{designmode}'s: \checkmark \\ Dark \texttt{designmode}'s: \checkmark} \\
      \hline
      \texttt{iota} series (DR21) & \makecell{Moffat FWHM$=0.57''$ and $\beta = 1.66$ \\ \quad at APO (Bright \texttt{designmode}'s),\\Moffat FWHM$=0.70''$ and $\beta = 1.87$ \\ \quad  at APO (Dark \texttt{designmode}'s),\\Moffat FWHM$=0.69''$ and $\beta = 1.78$ \\ \quad  at LCO (Bright \texttt{designmode}'s),\\Moffat FWHM$=0.62''$ and $\beta = 1.90$ \\ \quad  at LCO (Dark \texttt{designmode}'s)} & \nodata & Eq.~\ref{eq:wing} & \makecell{Bright \texttt{designmode}'s: \checkmark \\ Dark \texttt{designmode}'s: \checkmark} \\
      \hline
      \end{tabular}
  \label{tab:offset_versions}
\end{table*}

\subsubsection{Determining PSF Core}\label{sec:off_core}

In Section \ref{sec:wings}, we were able to determine a set of empirical relations for the PSF at large angular separations using archival eBOSS/BOSS data that will be important for bright star avoidance (Section \ref{sec:bright_avoid}). However, due to low number statistics, we were not able to constrain the ``core" (small angular separation) region of the PSF, as such close alignments had been deliberately avoided when designing BOSS/eBOSS plates. Additionally, because the PSF may depend on which corrector is used or other differences between plate-era observations at APO and current APO and LCO observations, it is crucial to examine this core region with SDSS-V data, rather than the archival data. Because of this, we constructed a set of test observations using the SDSS-V FPS before allowing intentional offsets from bright targets. 

\begin{figure*}[!t]
	\centering
	\includegraphics[width=0.49\textwidth]{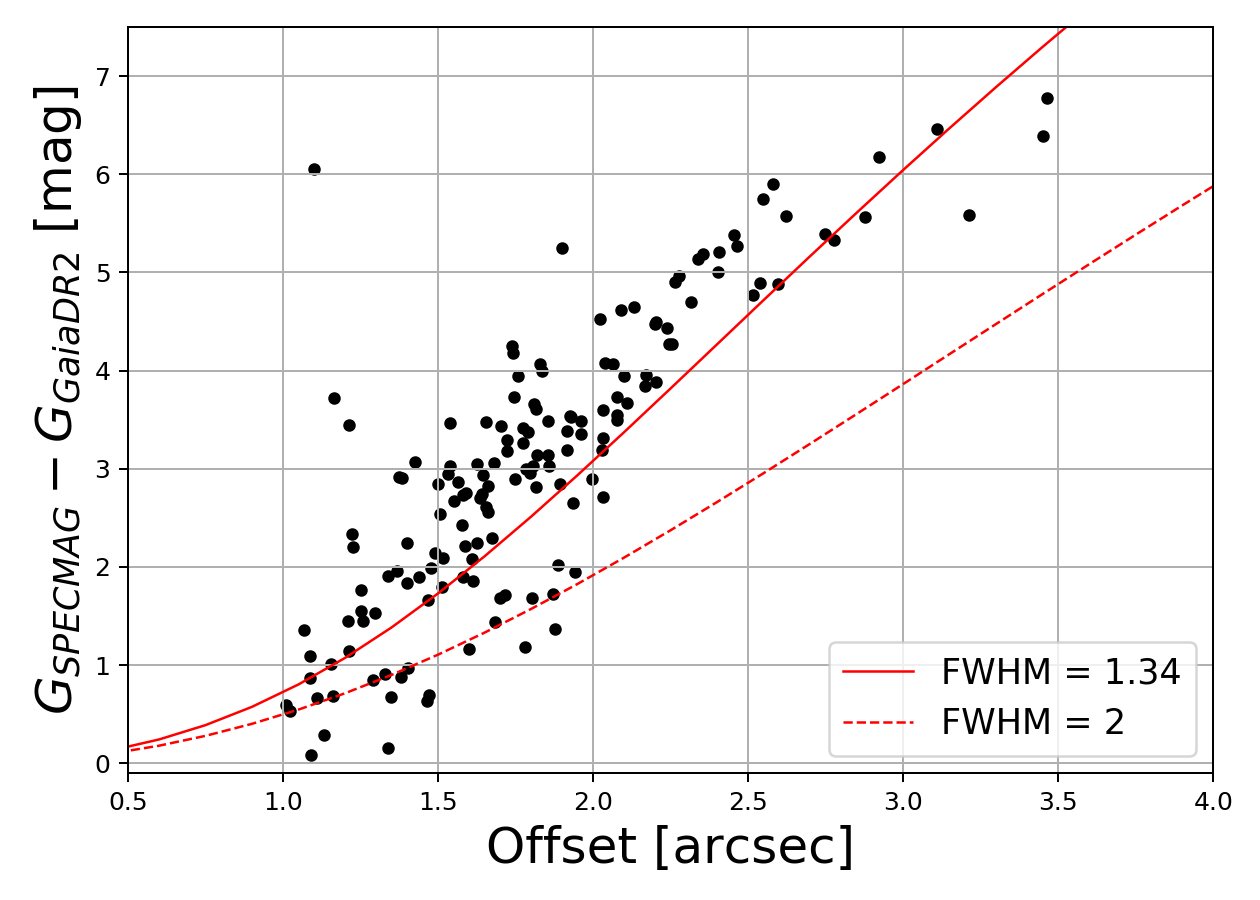}
    \includegraphics[width=0.49\textwidth]{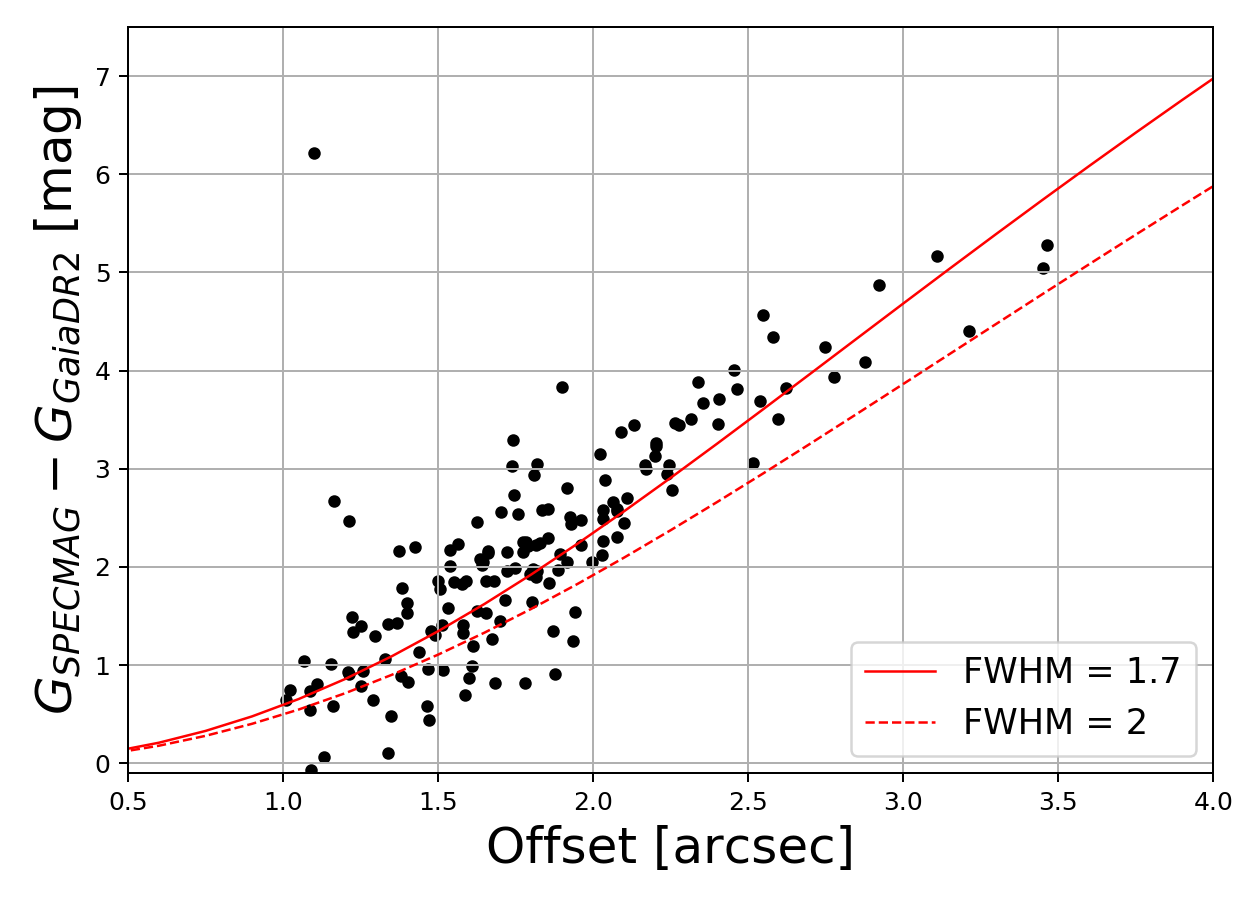}
    \caption{Magnitude loss versus fiber offset for an SDSS-V FPS test \textit{Design} at APO taken on MJD$=$59755 (left panel) and MJD$=$59760 (right panel). In both plots the lines shows the model prediction for the magnitude loss as a function of offset, where the model is a Moffat profile convolved with the circular fiber aperture. The dashed line in both plots shows a Moffat profile with FWHM$=$2$''$ and the solid lines shows a Moffat profile with a FWHM at the time of observation; FWHM$=$1.34$''$ for MJD$=$59755 (left panel) and FWHM$=$\edit1{1.7}$''$" for MJD$=$59760 (right panel).}
	\label{fig:moffat_test}
\end{figure*}

Figure \ref{fig:moffat_test} shows the results of such a test at APO. Here we created a \textit{Design} where we offset fibers from bright stars by some set number of arcseconds. These offsets were in a range such that we could probe the inner core of the PSF for the BOSS spectrograph. We repeated the observations for two separate nights so we could look at the magnitude loss as a function of fiber offset in different observing conditions. These results are shown as the data points in Figure \ref{fig:moffat_test}, were we see a monotonic increase in the magnitude loss (e.g., how much dimmer the magnitude from the spectra is than what is expected from \textit{Gaia}) as a function of the offset. To model this magnitude loss, we use a Moffat profile convolved with the circular fiber aperture. For a Moffat profile with a FWHM of 2$''$, we calculate the modeled magnitude loss shown as the dashed line in both plots. If we use a seeing typical for the night of observation (FWHM$=1.34''$ for MJD$=$59755, left panel, and FWHM$=1.7''$ for MJD$=$59760, right panel), we measure a magnitude loss that agrees well with the model, though the measurements are slightly larger than expected at larger offsets. Overall, based on subsequent tests it seemed that for APO the core of the PSF could be well described by a Moffat profile with FWHM$=$1.7$''$ and $\beta = 5$.

\begin{figure}[!t]
	\centering
	\includegraphics[width=0.45\textwidth]{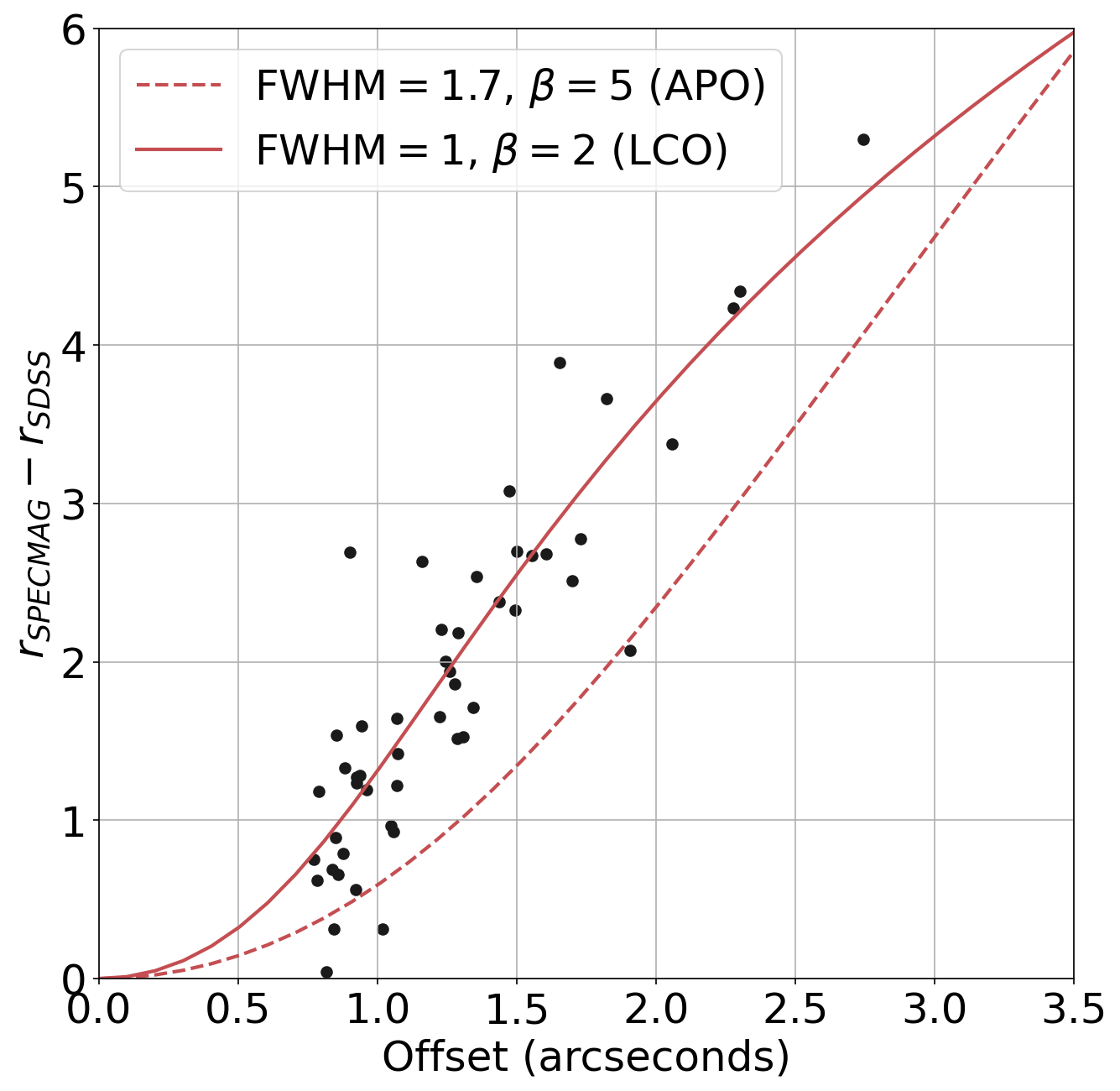}
	\caption{Magnitude loss versus fiber offset for an SDSS-V FPS test \textit{Design} at LCO taken on MJD$=$60124 (data points). In the plot, the lines shows the model prediction for the magnitude loss as a function of offset for APO (dashed line) and LCO (solid line), where the model is a Moffat profile convolved with the circular fiber aperture. The dashed line shows the APO offset model with FWHM$=1.7$"and $\beta = 5$, and the solid line shows the LCO offset model with FWHM$=1$"and $\beta = 2$.}
	\label{fig:moffat_test_LCO}
\end{figure}

Similar tests had to be performed for LCO. This is due to the different typical seeing conditions and plate scale of the focal plane system compared to APO. Similar to APO, we complete test observations at LCO and determined the magnitude loss versus fiber offset from the targets. Figure \ref{fig:moffat_test_LCO} shows the magnitude loss versus fiber offset for one of these LCO tests performed on MJD$=$60124. In this plot, we show the Moffat profile that models the PSF core for APO as the dashed line (see the corresponding fit  to the APO data in the right panel of Figure \ref{fig:moffat_test}). Here we see that this model greatly underestimates the magnitude loss compared to the observations. Instead, a Moffat profile with FWHM$=1''$ and $\beta = 2$ better matches these observations at LCO and is what we will use to model the PSF for this observatory.

\begin{figure*}[!t]
	\centering
	\includegraphics[width=\textwidth]{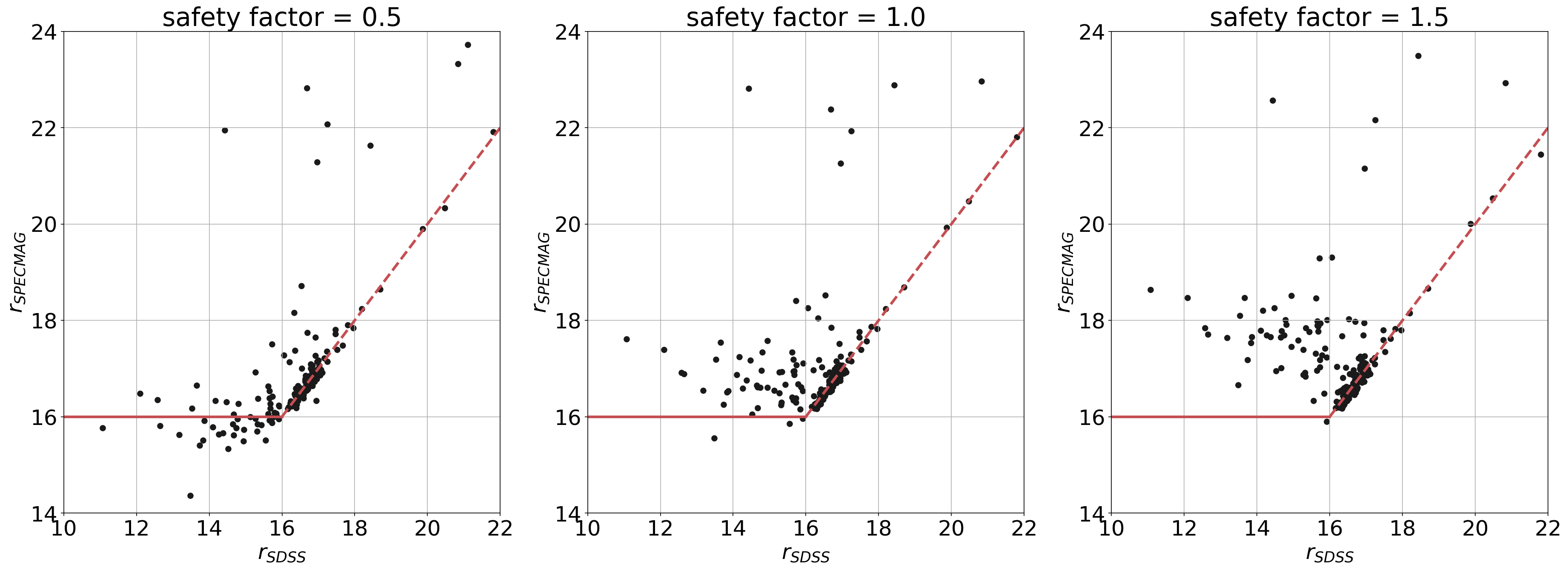}
	\caption{SDSS r-band magnitude estimated from the spectrum versus true r-band magnitude offset test at LCO taken on MJD$=$60124. Each panel shows the offsets calculated with a different value of the safety factor; a constant value added to the desired magnitude loss when calculating the offset of the fiber. In each plot, the solid red line shows the magnitude limit for the observation ($r=16$ mag) and the dashed line is a one-to-one line for targets that were not offset.}
	\label{fig:safety_factor_LCO}
\end{figure*}

In all of the above test observations, there is significant scatter around the Moffat profile used to model the offsets. \edit1{This could be due to differences in seeing and transparency at the time of observation. As for normal survey operations we plan the \textit{Designs} well ahead of time and as such conditions are stochastic, we do account for these at the time of fiber placement. As a result though, this scatter} could cause issues if \edit1{the conditions change such that the} magnitude loss is overestimated \edit1{and} the flux down the fiber is greater than expected. Because of this, when implementing offsets we utilize a ``safety factor." This safety factor is a constant value added to the desired magnitude loss when calculating the offset of the fiber. With an appropriate safety factor, even if there is some scatter around the function all offset targets should still be dimmer than the magnitude limit for the observation. Figure \ref{fig:safety_factor_LCO} shows such a test at LCO during dark time on MJD$=$60124, where each panel shows the SDSS r-band magnitude estimated from the spectrum versus true r-band magnitude. The red solid line shows the magnitude limit for the dark time observation, which is 
$r_{\rm SDSS} = 16$. For the offset targets ($r_{\rm SDSS} < 16$), a safety factor of $0.5$ results in many targets having magnitudes brighter than the limit for the observation. At a safety factor of 1, almost all of the offset targets are fainter than the magnitude limit. When we increase the safety factor to 1.5, the offset targets become excessively faint, which could yield unusable results in this observing mode. Because this test was performed during dark time, where we are most concerned about contamination from on chip neighbors, we choose to use a safety factor of 1 for the offsetting in dark time. We also performed similar tests during bright time, and we concluded a safety factor of 0.5 was adequate for these \texttt{designmode}'s, because we are less concerned about contamination and are therefore more willing to allow some observations be brighter than the magnitude limit to increase the signal-to-noise of the spectra.

\subsubsection{Implementation of Fiber Offset}

With the PSF defined in the outer regions (eqs. \ref{eq:trans}$-$\ref{eq:wing}) and in the inner core (the Moffat profiles described in Section \ref{sec:off_core}), we can implement offsets in the following manner. For all of these functional forms for the PSF, they are described as a magnitude loss as a function of distance from the center of the target. Because we normally want an offset for a desired magnitude loss, we invert all formulations, where this is done analytically for eqs. \ref{eq:trans}$-$\ref{eq:wing} and done numerically via.~linear interpolation for the Moffat profile.

For each target, the desired magnitude loss is the difference between the magnitude limit for the \textit{Design} (\texttt{bright\_limit\_targets\_min}; Table \ref{tab:targetdb-designmode-values}) plus the safety factor (0.5 for Bright Time and 1 for dark time data collection scenarios) and the magnitude of the target. The offset that should give this desired magnitude loss is calculated for each regime (core, transition region and PSF wings) and the offset is the maximum value from these various functional forms. This process is repeated for each optical bandpass ($g, r, i, z, B_P, G, R_P$) with a magnitude limit for BOSS or each infrared bandpass ($J, H, K$) for APOGEE, and the final offset is the maximum value from the various photometric bands. In the FPS, offsets are always applied in the positive Right Ascension direction. Ideally, we would always make the offset perpendicular to the parallactic angle to minimize chromatic effects associated with atmospheric differential refraction, but that would lead to unexpected collisions at observation time; choosing an offset in Right Ascension makes the offset more predictable while often being close to the desired direction. Additionally, offsets are prohibited for targets that are brighter than $G=6$ in Bright Time and $G=13$ magnitudes for dark time data collection scenarios for BOSS, and $H=1$ in all scenarios for APOGEE. So, such targets will not be observed during SDSS-V. \edit1{We note that when calculating offsets for targets, we do not strictly check if the offset location places the fiber on/near a nearby source. The exception is for the avoidance of bright neighbors (Section \ref{sec:bright_avoid}), which ensures that the placement of a fiber is not too close to a star brighter than the magnitude limit for the \textit{Design}. This means that in very crowded regions, there is a chance the fiber could be offset from one source to another.}

\edit1{The above} process is implemented at the survey planning stage to ensure robots can reach targets, but in practice offsets are calculated on-the-fly at the time of observation. The above is implemented within \texttt{coordio}\footnote{\url{https://github.com/sdss/coordio}}, the SDSS-V software product that handles all coordinate conversions. It should be noted that during the periods covered by the current public data releases of SDSS-V \citep[DR18 and the soon to be released DR19;][]{sdssDR18, sdssdr19}, offsetting was not implemented (i.e. objects brighter than the nominal limits were not targeted). Future data releases will include survey plans with offsetting such that these brighter targets will be included. This is summarized in Table \ref{tab:offset_versions}.

\subsubsection{Updates to the PSF Profile Over the Survey}

One of the convenient aspects of the implementation of the offset capability is that all offsets are calculated on-the-fly at the time of observation. This means that we can update how the offsets are calculated without creating an entirely new survey plan. Because of this, we have updated the definition of the PSF throughout the survey as our understanding of the FPS system improves. We do note that these updates have, to date, corresponded with the release of an updated survey plan. This is summarized in Table \ref{tab:offset_versions}. As can be seen in this table, initially we did not perform any offsetting and only implemented bright star avoidance (discussed in the next section) using our initial approximation of the PSF (described in Section \ref{sec:wings}). After performing more commissioning tests, we were able to better describe the core of the PSF (Section \ref{sec:off_core}) with a Moffat profile and begin intentionally offsetting from bright targets, which was accounted for in the survey plan to be released in DR20. We do note that we only initially allowed offsets during Bright Time, as we were more concerned about on chip contamination during Dark \texttt{designmode}'s.

\begin{figure*}[!t]
	\centering
	\includegraphics[width=\textwidth]{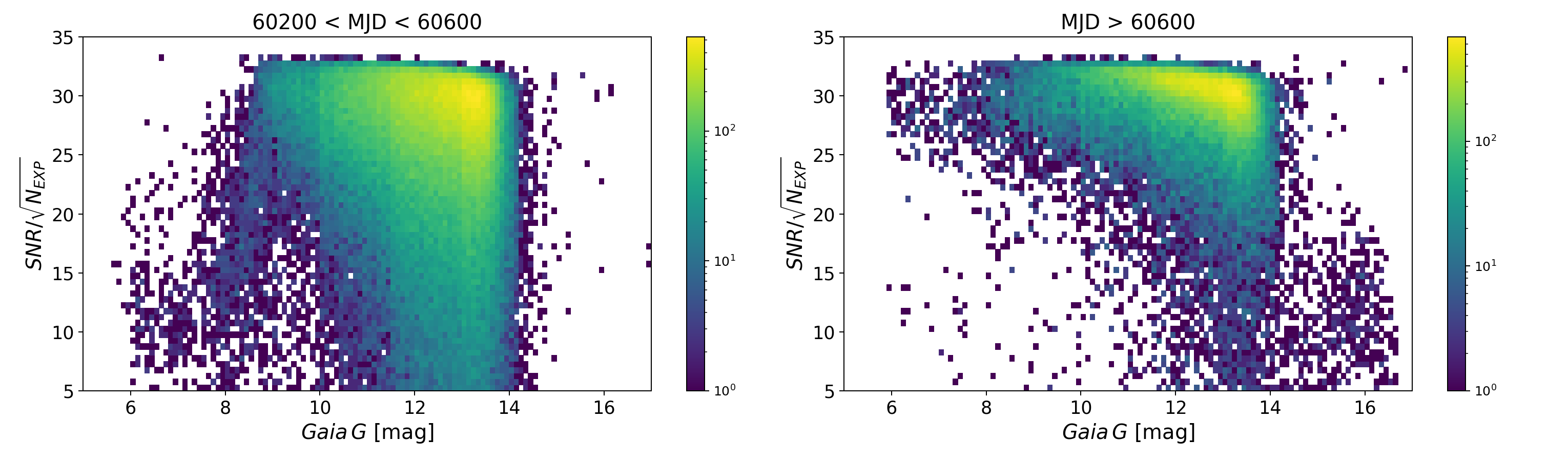}
	\caption{The \edit1{median} $SNR\sqrt{N_{exp}}$ vs.~Gaia G magnitude for offset targets in the eta series of the survey plan (left panel) and the theta series (right panel). During the eta series, very bright offset targets were receiving insufficient signal, indicating the offset function was too conservative. The function is updated for the theta series, such that the majority of offset targets now acquire a similar SNR.}
	\label{fig:update_offsets}
\end{figure*}

After observing in this mode for some time \edit1{with the \texttt{eta} series of the survey plan}, we noticed that our initial fits of the Moffat profile did not result in the expected flux down the fiber for targets brighter than $G\sim9$ mag. This is shown in the left panel of Figure \ref{fig:update_offsets}, where we see that at the bright end the SNR is much less than for the fainter stars that were offset. This indicates that our offset function is too conservative for the largest offsets. Additionally, there is a large amount of scatter in the SNR for all offset targets. Because of this, we used the predicted vs.~observed flux for the stars offset during science operations to recalibrate the Moffat profiles. Here we found that the Moffat profile was best described when independently considering both the observatory and the lunation during the observation. The latter is driven by the fact that we typically observe dark time designs at smaller airmasses than in bright time. Additionally, we found that the Moffat profile could be used in the ``transition" region, meaning our initial approximation from eq.~\ref{eq:trans} was no longer needed. The right panel of Figure \ref{fig:update_offsets} shows the SNR for offsets with these updated functions \edit1{during the \texttt{theta} series of the survey plan}, where we now see that at the bright end targets are getting sufficient signal and the scatter for all offset targets is greatly reduced, indicating we are better modeling the PSF. The new parameters for these Moffat profiles are summarized in Table \ref{tab:offset_versions}. Also, because we became more confident in our modeling of the PSF for offsets, we then allowed offsetting for \textit{Designs} with Dark \texttt{designmode}'s. \edit1{Similar updates were made before the creation of the following \texttt{iota} series of the survey plan, as with the addition of more dark time data we noticed similar updates were needed to achieve sufficient SNR.} Targets offset using these profiles are part of versions of the survey plan that will partially be released during DR20, with the majority being released in DR21. The offset data included in DR19 will only be from the commissioning tests outlined in Section \ref{sec:off_core}.

As we have utilized the offset feature for only portions of the survey and have changed how it is implemented, this does have an effect on the SDSS-V selection function. For example, during the \texttt{eta} series of the survey plan, we allowed offsetting during bright time, but not dark time. This means a program attempting to observe bright stars in both bright and dark led \textit{Fields} would only get assignments in the bright ones, leading to a noticeable selection effect. Once we transitioned to the \texttt{theta} series though, dark time offsets were allowed and the distribution of stars in that program would better match the survey footprint. These subtleties must be considered when defining the SDSS-V selection function.

\subsection{Implementation of Bright Star Avoidance}\label{sec:bright_avoid} 

In SDSS-V, various limits on the number of skies, standards, magnitude limits of objects, etc. are utilized in different data collection scenarios with specific \texttt{designmode} parameters to unsure the science requirements of the observation are met. One important aspect of this is the minimum magnitude of an object in a given \textit{Design}. These magnitude limits are in place to ensure that the faintest science targets are not adversely affected by contamination from bright on-chip neighbors (see Section \ref{sec:mag_limits}). While no stars exceeding the bright limits are assigned to fibers in \texttt{robostrategy}, fibers could still be placed sufficiently close to a bright star such that the flux entering the fiber exceeds the bright limit for the \texttt{designmode} of the \textit{Design}; we must also perform a ``bright neighbor check" to identify fibers that meet this criteria.

For the bright neighbor check, all objects brighter than the minimum magnitude limit for the \texttt{designmode} within the field of view are identified from various catalogs. For BOSS, objects from the union of \textit{Gaia} DR2 or DR3 (depending on the targeting version) and Tycho-2 are used (favoring the former when a star is contained in both catalogs) . 
To place the \textit{Gaia} and Tycho-2 objects on a common system, \textit{Gaia} G-band magnitudes for the Tycho-2 objects are estimated as follows. For Tycho-2 objects that have both $B_T$ and $V_T$ measurements, the following relation from \cite{evans2018} is used:
\begin{equation}
\begin{split}
    G =  V_T - 0.02051 & - 0.2706 \times (B_T - V_T)\\ 
                       & + 0.03394 \times (B_T - V_T)^2\\
                       & - 0.05937 \times (B_T - V_T)^3
\end{split}
\end{equation}
and for Tycho-2 objects with only $V_T$ measurements, we use:
\begin{equation}
    G = V_T - 1
\end{equation}
For APOGEE, objects from 2MASS are used for the bright neighbor check and the standard 2MASS H-band magnitudes are used in the subsequent check.

With this collection of bright objects, we find the ``exclusion radius" around each one. Here, the exclusion radius is defined as the radial distance from a bright source within which the excess flux from the bright object exceeds the minimum magnitude limit for the \texttt{designmode} of the \textit{Design}. This estimation of the excess flux as a function of radius from the object matches the functional forms used for the various versions of the survey plan summarized in Table \ref{tab:offset_versions}. It should be noted that exclusion radii are calculated without the safety factor that is used for offsetting. With these exclusion radii, at the \textit{Design} creation stage \texttt{robostrategy} ensures that no fibers are assigned within this radius and at the validation stage \texttt{mugatu} has routines to ensure a \textit{Design} meets this criteria.

In the FPS, there are robots that have both an APOGEE and BOSS fiber. It could then be possible that with this bright star avoidance, it can be impossible to observe bright APOGEE targets during dark time. \edit1{This could occur} if the excess flux from the bright APOGEE source \edit1{is} large enough \edit1{such that} the BOSS fiber \edit1{is} within its exclusion radius. In practice, this does not occur. In the design of the FPS \citep{FPS}, the distance between these fibers on the same robot was set to be larger than the PSF wings of a bright star centered on one fiber. This distance ends up being $\sim4.7"$ at APO and $\sim8.2"$ at LCO. At these distances, the effect of one assigned bright star effecting the assignment of the adjacent fiber is negligible.

\section{Implementing Constraints in Design Creation}\label{sec:implement}

The survey planning software, \texttt{robostrategy} \citep{blanton2025}, creates all of the \textit{Designs} to respect these \texttt{obsmode} and \texttt{designmode} constraints. The full details about how and the order in which these constraints are considered can be found in \citet{blanton2025}, but a brief summary in given here. 

\texttt{robostrategy} decides the cadence to observe for each \textit{Field} and how to 
assign targets to fibers for each \textit{Design} in the \textit{Field}. The \textit{Field} cadences specify
the desired timing of the observations of each \textit{Design}, and the data collection 
scenario. They are chosen to maximize a metric expressing the science value of 
the observations, under constraints on how much dark and bright observing time 
there is available. The bright-time associated cadences all use the Bright Time
scenario. The dark-time associated cadences are set based on the dominant program, as outlined in Section \ref{sec:obs_mode}. RM \textit{Fields} use the
Dark RM scenario, \textit{Fields} associated with the All-Quasar Multi-Epoch Spectroscopy
(AQMES) program use Dark Monitoring, \textit{Fields} led by the  SPectroscopic 
IDentfication of ERosita Sources (SPIDERS) program use Dark Faint, and all other
dark-time \textit{Fields} (typically dark time at low Galactic latitude, which allows 
observations of white dwarfs) use Dark Plane.

For each \textit{Field}, once its cadence is known, the first thing \texttt{robostrategy} 
does is to determine what the maximum number of calibration targets it can observe 
in the \textit{Field} and (for APOGEE standards) in each focal plane zone; if, as happens 
in rare cases, any of these numbers is lower than the requirement, the requirement 
is loosened for the \textit{Field}. When \texttt{robostrategy} assigns targets, it assigns 
robots to targets in order of priority. As it does so, it checks whether fibers 
can still reach the required number of calibration targets in the \textit{Field} and 
(for APOGEE standards) in each focal plane zone, allowing for swaps between 
fiber assignments between science and calibration targets if necessary. During 
each target assignment, \texttt{robostrategy} checks if an offset is necessary 
and whether the bright star exclusion radius criteria are respected.

When assigning targets to fibers, \texttt{robostrategy} must also respect the physical constraints of the FPS hardware. Specifically, each target-to-fiber assignment must (1) land within the focal plane patrol zone of its intended robot and (2) not create collisions with other robots given their respective target assignments. \texttt{robostrategy} relies upon the \texttt{kaiju} software product \citep{Sayres2021} to evaluate these (and other) geometrical constraints during target assignment optimization. Each \textit{Design} is created assuming a nominal LST for observation such that all \textit{Designs} fit within the time constraints of a five year survey. In the nightly scheduling of observations, the LST of observation for a \textit{Design} may differ from its \texttt{robostrategy} prescribed LST. Subtle time and temperature varying effects at the telescope, such as plate scale and differential atmospheric refraction, will perturb a target’s location in the focal plane. These perturbations will occasionally move an object outside of the patrol zone of its intended robot or create a target assignment collision with another robot. Additionally, throughout the survey robots occasionally malfunction and become disabled. Over the first two years of the survey, a median of 3\% of robots in a \textit{Design} could not make it to their assigned target at time of observation due to such issues.

All of the parameters listed in Section \ref{sec:desmode} are also validated by \texttt{mugatu}\footnote{\url{https://github.com/sdss/mugatu}}, even the ones that are not specifically used by \texttt{robostrategy} to constrain assignments during the survey planning. Additionally, \texttt{mugatu} utilizes \texttt{kaiju} to ensure that all assignments are valid (i.e.~robots can reach assignments and not cause collisions). The validation results are saved and tracked for each iteration of the survey plan, since in a small number of cases fail one or more \texttt{designmode} criteria. For example, this could be due to especially sparse regions on the sky where there are insufficient telluric standards to either meet the minimum requirement or to be well distributed across the FOV. Apart from the parameters that are not considered by \texttt{robostrategy} (e.g., FOV distributions of BOSS standards and skies, and APOGEE skies), these failures happen in a very small percentage of \textit{Designs} ($\sim 0.1-5\%$ of \textit{Designs}, depending on the parameter and observing mode).

An example of these results in shown in Table \ref{tab:zeta_3_valid}, which shows the percentage of \textit{Designs} that pass a \texttt{designmode} criteria for \texttt{zeta-3}. This is the survey plan version released with DR19 \citep{sdssdr19}, though we do note that only APO data will be part of that data release. Here, with the exception of the \texttt{FOV} metrics, typically $>99\%$ of \textit{Designs} meet the \texttt{designmode} criteria. \edit1{Besides the \texttt{FOV} metrics, these failure cases are in the minimum required skies and standards for a \textit{Design}. These \textit{Designs} are simply cases where there are not enough available skies and/or standards in the \textit{Field} to satisfy the requirement. In these instances, we still create and observe the \textit{Design} though, as this is the best set of assignments we can create for this \textit{Field} given the targeting data. The small number of \textit{Designs} that are insufficient in skies and/or standards are tracked all the way to the public data release, where they can be identified in the \texttt{mos\_design\_mode\_check\_results} table.}

\begin{figure*}[!t]
	\centering
	\includegraphics[width=\textwidth]{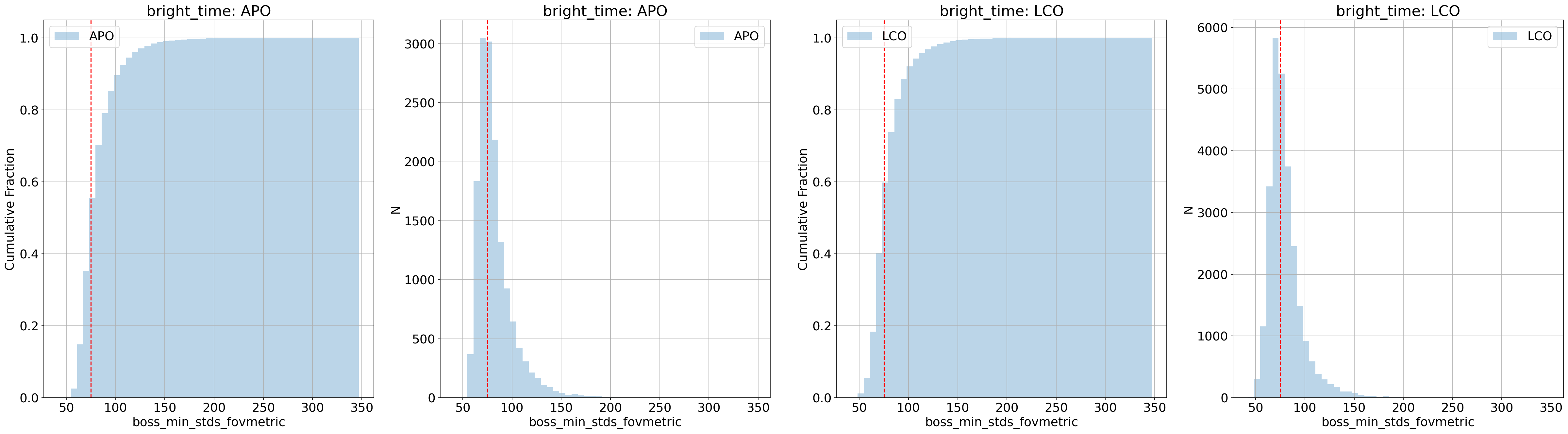}
	\caption{Cumulative distribution and histogram for $r_p$ in \texttt{stds\_FOV} metric for BOSS all Bright Time \textit{Designs} in the \texttt{zeta-3} version of the survey plan. The red dashed line shows the $d$ value for the Bright Time \texttt{stds\_FOV} metric.}
	\label{fig:boss_fov_dist}
\end{figure*}

\edit1{For the other results,} there are a few subtleties to \edit1{discuss}. For Dark RM, this only includes three \textit{Fields}, where all \textit{Designs} in a \textit{Field} consist of the same fiber assignments. This can throw off these percentages. For example, for \texttt{skies\_FOV} the $r_p$ values are 78.2 mm, 81.5 mm and 98.6 mm. While nominally greater than $d$ for this metric, because of the number of skies the distribution is good enough for calibration (see Figure \ref{fig:fov_metric_boss}). Additionally, for the \texttt{stds\_FOV} criteria for all data collection scenarios, we of course see lower pass rates for BOSS as this is not constrained by \texttt{robostrategy} during the \textit{Design} creation process. This pass/fail metric is only a snapshot though. Along with these statistics, \texttt{mugatu} also creates other diagnostic plots during the validation process. This includes sky plots (Figure \ref{fig:apogee_fov_valid}) and the distributions of the metrics (Figure \ref{fig:boss_fov_dist}). As can be seen from Figure \ref{fig:boss_fov_dist}, $\sim90\%$ of \textit{Designs} have a distribution of standards similar to what was shown in Figure \ref{fig:fov_metric_boss} during Bright Time. This type of analysis has been used to guide us throughout the survey to efficiently examine the quality of the \textit{Designs} in the proposed survey plans.

One aspect of the validation process from \texttt{mugatu} that is important to highlight is its efficiency. With the recent version 2.5.0 of \texttt{mugatu}, the entire survey ($\sim42,000$ \textit{Designs}) can be validated within $\sim 95$ CPU hours. As stated above, there is then some human processing of the statistics that need to take place, which is largely handled through examining sets of summary plots and statistics (like Table \ref{tab:zeta_3_valid}), but because of the framework described here this human processing time is \textit{much} less than in previous versions of SDSS. This greatly differs from plate operations where each plate design has to be visually inspected by one of the collaboration members. This would be infeasible for a survey of the size of SDSS-V. With this efficiency, we also now have the power to run many versions of the survey plan and be able to quickly validate them.

\begin{table*}[!t]
\footnotesize
  \centering
  \caption{\texttt{zeta-3} validation results from \texttt{mugatu}. For each data collection scenario, observatory and instrument, the percentage of \textit{Designs} that passed the \texttt{designmode} metric are shown.} 
  \begin{tabular}{|l | l |l |c |c |c |c |c |}
       \cline{4-8}
      \multicolumn{3}{c|}{} & \multicolumn{5}{c|}{\textit{\textbf{Data Collection Scenario}}} \\
       \hline
       \texttt{designmode} & \textbf{Observatory} & \textbf{Instrument} & \textbf{Bright Time} & \textbf{Dark Faint} & \textbf{Dark Monit.} & \textbf{Dark Plane} & \textbf{Dark RM} \\
       \hline
       \hline
       \multirow{4}*{\texttt{skies\_min}} & \multirow{2}*{APO} & BOSS & 99.97\% & 100\% & 99.98\% & 100\% & 100\% \\
       \cline{3-8}
                                          &                    & APOGEE & 99.42\% & 98.44\% & 97.54\% & 100\% & 100\% \\
       \cline{2-8}
                                          & \multirow{2}*{LCO} & BOSS & 99.99\% & 100\% & \nodata & 100\% & \nodata \\
       \cline{3-8}
                                          &                    & APOGEE & 99.91\% & 100\% & \nodata & 100\% & \nodata \\
       \hline
       \hline
       \multirow{4}*{\texttt{skies\_FOV}} & \multirow{2}*{APO} & BOSS & 31.09\% & 79.40\% & 35.14\% & 36.42\% & 0\% \\
       \cline{3-8}
                                          &                    & APOGEE & 21.87\% & 40.91\% & 50.97\% & 25.74\% & 100\% \\
       \cline{2-8}
                                          & \multirow{2}*{LCO} & BOSS & 39.13\% & 82.22\% & \nodata & 43.47\% & \nodata \\
       \cline{3-8}
                                          &                    & APOGEE & 22.21\% & 50.80\% & \nodata & 34.74\% & \nodata \\
       \hline
       \hline
       \multirow{4}*{\texttt{stds\_min}} & \multirow{2}*{APO} & BOSS & 99.99\% & 100\% & 100\% &100\% & 100\% \\
       \cline{3-8}
                                          &                    & APOGEE & 99.99\% & 99.72\% & 100\% & 100\% & 100\% \\
       \cline{2-8}
                                          & \multirow{2}*{LCO} & BOSS & 99.93\% & 100\% & \nodata & 99.31\% & \nodata \\
       \cline{3-8}
                                          &                    & APOGEE & 98.05\% & 97.24\% & \nodata & 94.91\% & \nodata \\
       \hline
       \hline
       \multirow{4}*{\texttt{stds\_FOV}} & \multirow{2}*{APO} & BOSS & 41.51\% & 70.24\% & 91.34\% &48.52\% & 100\% \\
       \cline{3-8}
                                          &                    & APOGEE & 97.50\% & 99.57\% & 99.57\% & 99.11\% & 100\% \\
       \cline{2-8}
                                          & \multirow{2}*{LCO} & BOSS & 45.86\% & 77.75\% & \nodata & 57.66\% & \nodata \\
       \cline{3-8}
                                          &                    & APOGEE & 99.05\% & 99.53\% & \nodata & 99.37\% & \nodata \\
       \hline
       \hline
       \multirow{4}*{\texttt{stds\_mags}} & \multirow{2}*{APO} & BOSS & 100\% & 100\% & 100\% &100\% & 100\% \\
       \cline{3-8}
                                          &                    & APOGEE & 100\% & 100\% & 100\% & 100\% & 100\% \\
       \cline{2-8}
                                          & \multirow{2}*{LCO} & BOSS & 100\% & 100\% & \nodata & 100\% & \nodata \\
       \cline{3-8}
                                          &                    & APOGEE & 100\% & 100\% & \nodata & 100\% & \nodata \\
       \hline
       \hline
       \multirow{4}*{\texttt{bright\_limit\_targets}} & \multirow{2}*{APO} & BOSS & 100\% & 100\% & 100\% &100\% & 100\% \\
       \cline{3-8}
                                          &                    & APOGEE & 100\% & 100\% & 100\% & 100\% & 100\% \\
       \cline{2-8}
                                          & \multirow{2}*{LCO} & BOSS & 100\% & 100\% & \nodata & 100\% & \nodata \\
       \cline{3-8}
                                          &                    & APOGEE & 100\% & 100\% & \nodata & 100\% & \nodata \\
       \hline
      \end{tabular}
  \label{tab:zeta_3_valid}
\end{table*}

Another important feature of this entire implementation is its utility in further testing of features throughout SDSS-V. This process is largely modular and can allow for singular \textit{Designs} to be made with a range of parameters for the \texttt{obsmode} and \texttt{designmode}. Additionally, \texttt{robostrategy} is set up such that \textit{Designs} can be created that do not constrain assignments based on certain \texttt{designmode} parameters. This allows for custom \textit{Designs} to be created for survey commissioning, which includes the testing of new features that will be implemented further in the survey. This is the mode in which \textit{Designs} were created for testing the offset capabilities of the FPS (Section \ref{sec:offsets}).

\section{Summary}\label{sec:summary}

The science goals of SDSS-V are ambitious in nature, as the SDSS-V MOS program is, in general, simultaneously observing a more heterogeneous mix of targets than previous SDSS iterations.. These goals are largely facilitated by the addition of the FPS, which is newly equipped with robot positioners and allows for quick changes in the configuration of targets between observations. With these additions comes increased complexity during the survey plan. This is especially due to the larger variety of targets, which makes it more difficult to create unique \textit{Designs} that respect the needs of all science programs. To accomplish this ambitious goal, we created a framework that allows for the algorithmic assignment of targets from a larger variety of programs, while still respecting the overall science outcomes needed for all targets.

This is accomplished by first defining a list of data collection scenarios. For each scenario, we lay out the prioritized science goals and the types of targets that are of most concern. This allows for us to prioritize faint objects in one \textit{Design}, accurate spectrophotometry in another, etc. For each data collection scenario, we then define a set of parameters that are used by the survey planning tool to ensure these goals are met. These are broadly broken up into \texttt{obsmode} and \texttt{designmode} parameters. Here, the \texttt{obsmode} parameters specify the sky conditions at the time of observation of the \textit{Design}, while the \texttt{designmode} parameters specify the spectrograph specific calibration requirements and fiber assignment restrictions. In this paper we largely focus on the definitions of the latter.

Here the \texttt{designmode} parameters are set to generally allow for the desired reduction outcomes for a data collection scenario. This is accomplished by:
\begin{itemize}
    \item Setting the number of calibrators to have the desired final data quality. This is a balance between absolute flux calibration and survey efficiency. When accurate spectrophotometry is needed, more calibrators are required. When this is not needed, less calibrators are required to increase survey speed.
    \item Restricting the distribution of calibrators across the FPS such that changes \edit1{in the effective throughput and sky flux across the field of view can be accounted for}.
    \item Setting magnitude limits on the assigned targets to limit contamination from on chip neighbors.
    \item Further limiting the position of fibers to avoid excess flux from bright sources that may cause unwanted contamination on the chip.
\end{itemize}
The above parameters are set through various tests that utilized both archival SDSS data and commissioning data from SDSS-V. With the parameters set, the various SDSS-V software products could then algorithmically restrict the assignment of targets to fibers both during survey planning and at the time of observations in the case of offsetting fibers. This was a huge paradigm shift from previous iterations of SDSS, as for SDSS-V this framework truly allowed for a singular, global definition of the observational constraints that could be applied and facilitate the planning of the entire survey.

This document finally serves two important purposes. First, the understanding of these constraints are crucial for detailed modeling of the SDSS-V selection function. These parameters first and foremost set the constraints on which targets can be assigned to a fiber throughout the length of the survey, so they must be accounted for when modeling its selection function. \edit1{Second}, we hope that this framework will serve as a model for future multiplexed, spectroscopic surveys. As our technology continues to improve and allows for increased survey speed from robotically controlled focal plane systems, similar frameworks will need to be utilized to ensure a variety of science programs can coexist and the goals of the various programs are respected. We hope that the work summarized here will be be of great aid to the community for both of these purposes.


\section*{Acknowledgment}

EZ would like to thank Hans-Walter Rix, who first suggested to deliberately offset fibres to allow observations of the brightest stars. 

Funding for the Sloan Digital Sky Survey V has been provided by the 
Alfred P. Sloan Foundation, the Heising-Simons Foundation, the National 
Science Foundation, and the Participating Institutions. SDSS 
acknowledges support and resources from the Center for High-Performance 
Computing at the University of Utah. SDSS telescopes are located 
at Apache Point Observatory, funded by the Astrophysical Research 
Consortium and operated by New Mexico State University, and at Las 
Campanas Observatory, operated by the Carnegie Institution for Science. The SDSS web site is \url{www.sdss.org}.

SDSS is managed by the Astrophysical Research Consortium for the 
Participating Institutions of the SDSS Collaboration, including the 
Carnegie Institution for Science, Chilean National Time Allocation 
Committee (CNTAC) ratified researchers, Caltech, the Gotham 
Participation Group, Harvard University, Heidelberg University, The 
Flatiron Institute, The Johns Hopkins University, L'Ecole polytechnique 
f\'{e}d\'{e}rale de Lausanne (EPFL), Leibniz-Institut f\"{u}r 
Astrophysik Potsdam (AIP), Max-Planck-Institut f\"{u}r Astronomie (MPIA 
Heidelberg), Max-Planck-Institut f\"{u}r Extraterrestrische Physik 
(MPE), Nanjing University, National Astronomical Observatories of China 
(NAOC), New Mexico State University, The Ohio State University, 
Pennsylvania State University, Smithsonian Astrophysical Observatory, 
Space Telescope Science Institute (STScI), the Stellar Astrophysics 
Participation Group, Universidad Nacional Aut\'{o}noma de M\'{e}xico, 
University of Arizona, University of Colorado Boulder, University of 
Illinois at Urbana-Champaign, University of Toronto, University of Utah,
 University of Virginia, Yale University, and Yunnan University.

\software{Astropy \citep{astropy:2013, astropy:2018, astropy:2022}, Numpy \citep{numpy}, Matplotlib \citep{matplotlib}, Scipy \citep{scipy}}

%






\bibliography{determing_design_reqs}{}
\bibliographystyle{aasjournal}



\end{document}